# Electrohydrodynamic emitters of ion beams

V.G. Dudnikov

Muons, Inc, 552 North Batavia Ave. Batavia IL 60510.

ABSTRACT

Physical processes that determine the generation of ion beams with a high emission current density in electrohydrodynamic emitters are considered. Electrohydrodynamic effects, which are manifested in the features of ion emission, and the kinetics of ion interaction in high-density beams are discussed. The factors that determine the size of the emission zone, stability of emission at high and low currents, cluster generation, increase in energy dispersion, and decrease in beam brightness are analyzed. The problems of the practical maintenance of the stable functioning of the EHD emitters are considered. The practical applications of EHD emitters are considered.

## INTRODUCTION

In literature on the microtechnology (mainly of technologies of microelectronics), published in recent Years [1,2,3,4,5],became mandatory the section devoted to elektrohydrodynamic (liquid metal) ion source and systems with fine focused ion beam for micromachining in their basis: submicron lithography implantography, microanalysis. While up to now the role of EHD source in the production of microproducts is not to high, it is considered as important method of elements of novel technologies. It is assumed that in the combination with a raster tunneling microscope EHD source can start founding members of first nascent nanotechnology [6,7]. Emission characteristics of EHD sources is so high, the dating of him so much impression all ho familiar with a typical characteristics of other ion sources. In the EHD source by very simple means ensured a generation of stable stationary ion beam generation with emission current density of hundreds million A / $cm^2$. Like the result was not reached by development field electron emitters. At a typically emission ion current density at another sources less than ~1 A / $cm^2$ these results were absolutely fantastic. That to happen, during about a dozen years EHD sources development was not mentioned in publications in Russian. This activities is cultivated in foreign groups working field emission ion Muller projectors and in firms linked to manufacture technological equipment for microelectronics fabrication. Interest for this problem of the Soviet specialists of ion source was not manifested itself almost for about 1983. On All-union seminar of intense ion source and the beam in Kiev a question on a liquid metal sources in the first time was raised by I. M. Roife in May 1983. In his report was discussed results of foreign publications. Many party in workshop has shown big interest to these discussions. Little later in Journal "Physics Uspekchi" was published a review of M.D. Gabovich "Liquid-metal ion emitters" [8], witch contributed to familiarization of broad public to this problematic. In future in Kiev Seminars of intense ion sources and beams was discussed almost all soviet works on EHD emitters, published letter in periodic issues. Study of EHD emitters was started in many laboratories in the USSR, but really Interesting results was reached only little. The studies of lest years have changed existing representation on many features of functioning of EHD emitters. More well-established representations on EHD-emitters summation in the reviews [9,10,11]. Technologically applications of EHD emitters are devoted special reviews [3,4,5,12]. A review of the works on EHD emitters performed in the Soviet Union until 1991 are presented in [13]. A review of ion sources for ion microlithography is presented in [14]. In present work main attention is concentrated on the features of physical processes in EHD emitters, that was revealed in recent time and was weak reflected in literature. Researched process have an angstrom sizes and a superstronge electric fields, thu on should eliminate direct observation. We needs use indirect method registration and far extrapolation thu increase role physical model and the requirements to validity.

Naturally, the proposed materials should not be regarded as completing the development of this area. There are still many unresolved issues and wide scope for activity both in researching physical processes and in perfecting new unique legs of equipment based on EHD emitters.

## PHYSICAL FEATURES OF FUNCTIONIN EHD EMITTERS

The generation of ion beams in EHD emitters is largely determined by the behavior of the melt film on the electrodes in a strong electric field, i.e. electrohydrodynamics of such systems. With an increase in the electric field strength, field emission arises on the negative electrode, and with a positive polarity in a field with $E > 10^8$ V / cm, it is possible to tear out positive ions and intense field evaporation of the substance begins. This process was discovered by Müller in 1941 [15].

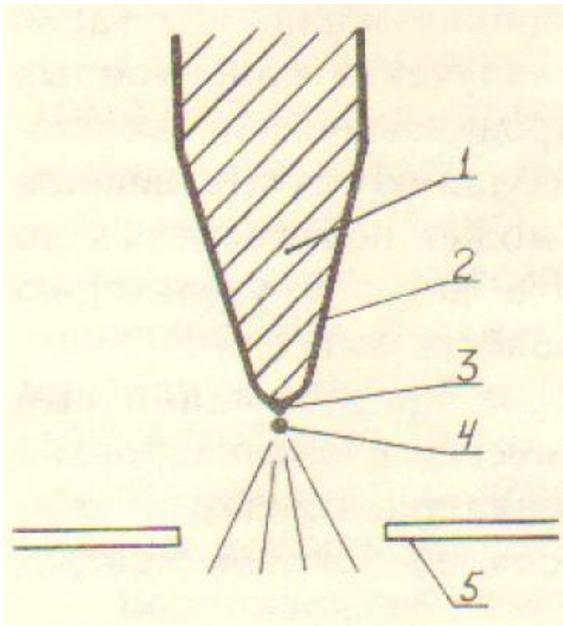

Fig. 1. Schematic representation of a working EHD emitter.

1-needle, 2-film of melt, 3-emitting sharpening, 4-region of luminescence, 5-extractor.

For a long time, EHD emitters were developed by space technology centers as small thrust engines for spacecraft. We used emitters in the form of a capillary, from the open end of which the molten material was atomized by an electric field in the form of charged microdrops - clusters [16]. A similar process has long been used for spraying paints. In 1969 (F. Machoney et al. "Electrohydrodynamic Ion Sources" [17]) confirmation of the emission of atomic ions from capillaries with a small outlet aperture d ~ 0.04 mm was obtained. In this paper, arguments are presented in favor of ion generation due to field evaporation and a rather extensive bibliography on EHD emitters is used. True, a very modest estimate of j ~ 30 A / cm² is given for the emission current density, and the assumption that the current density may be $j > 10^4$ A / cm$^2$ seemed too bold to the authors. It should be noted that the fabrication of thin metal capillaries is rather difficult. However, even with a small capillary diameter, the emitting tip of the liquid can move along the surface of the protruding drop, which leads to an increase in the effective size of the emitter and the instability of the emission.

The design of the EHD source, studied by R. Clampitt and D.K Jefferies., turned out to be successful. [18], in which the emission region is located on the tip of the needle extended through the open end of the capillary. The melt flows to the emission zone along the lateral surface of the needle. A schematic representation of the working EHD emitter is presented in Fig. 1. After that, various designs of EHD sources were developed, including very simple and very complex ones. Some examples are shown in Fig. 2 of [19]. A source with a needle made of porous tungsten impregnated with a working substance [20], gap sources, in which emitting peaks are located along a slit 1 μm wide and 8 cm long [21, 19], and sources in which the needle tip is located at a distance of ~ 10 μm from the target [7]. The latter, in terms of their design and control principle, are very similar to a scanning tunneling microscope and can be used to process a target, which at the same time is the source's extracting electrode. Most often EHD-emitters consists

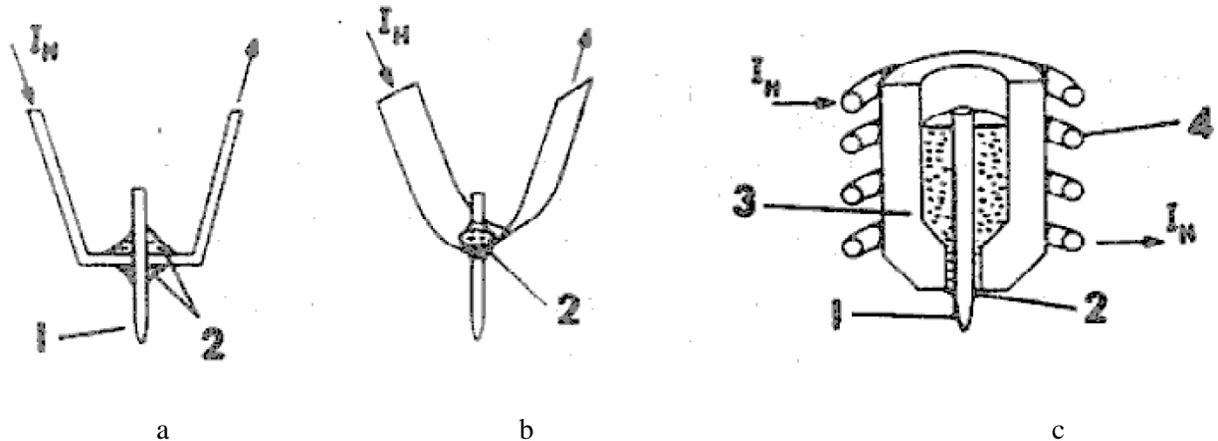

a                          b                          c

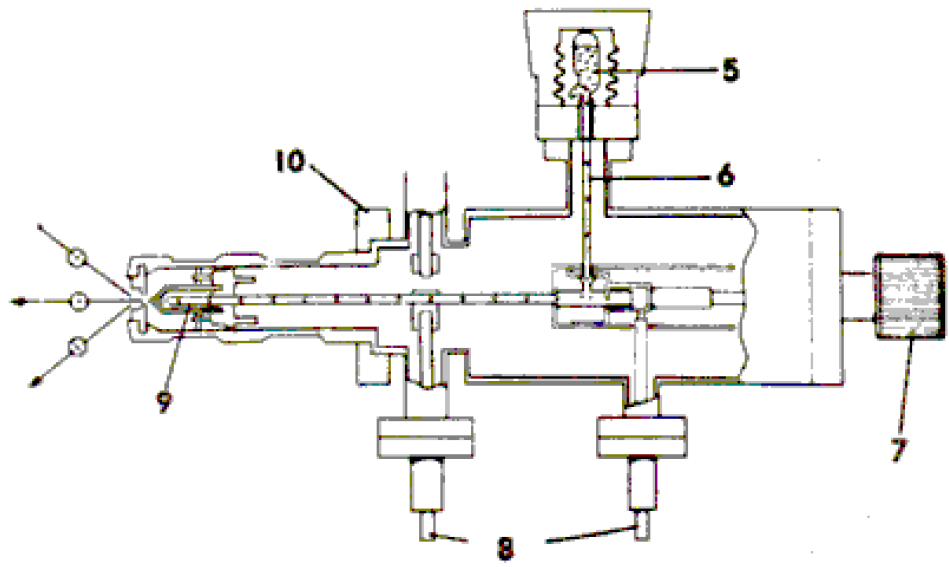

d

Fig. 2. Examples of constructions of EHD emitters. a, b — emitters with a bow; c — container emitter; d— EHD emitter of cesium ions, 1 — needle, 2 — liquid's target, 3 — container, 4 — heater, 5 — cesium ampoule, b — tube for filling the reservoir with cesium, 7 — needle movement mechanism, 5 — voltage input, 9 - a reservoir with a needle, 10 - a vacuum flange.

from a needle with a pointed radius of 3–10 μm moistened with a melt film. The electric field is concentrated between the needle and the extractor with an opening for the beam. Ions are emitted from the self-sharpened portion of the melt to the tip of the needle. At the base of the needle is a supply of working substance, which flows along the lateral surface to the tip. The needle and working substance are maintained at the right temperature using a heater. Typical operating voltages are 5–15 kV, emission current 0.5–200 μA. Metals and alloys that can be melted without rapid evaporation and destruction of the EHD emitter structure are used as working substances.

. In connection with the exclusive use of metal melts, these sources are called liquid metal. On the production of high-brightness ion beams in the mode of electrohydrodynamic emission from melts of dielectrics VG Dudnikov and A.L. Shabalin was reported in 1985 [22]. EHD emission from dielectric melts, which significantly expands the assortment of ions obtained, was later mastered by other authors [23,24]. .The preparation of negative Cl-, J-, F- ions from CsCl, CsI, CsF melts is described in [25].

FLUID IN THE ELECTRIC FIELD

The shape of the surface of a liquid in an electric field was studied by Taylor in [26]. He showed that with a specially selected shape of the electrodes, there is an equilibrium configuration of the liquid in the form of a cone with a semi-angle at the apex $\alpha = 49.3$ °. On the surface of the cone with precisely such an angle, the electric field strength depends on the distance to the apex as $r^{1/2}$, and at a certain voltage U, a balance of electric and capillary pressures is provided along the entire surface. The work of Taylor [26] is usually considered the justification for the fact that, at a critical potential difference Ucr between the extractor and the liquid metal emitter, the latter takes the form of a cone with a certain angle at the vertex $2\alpha_o$ (Fig. 2). The condition of pressure equilibrium on the surface of a conducting fluid with surface tension $\gamma$ and with an electric field E on this surface has the form

$$\gamma \, (1 \, / \, \rho_1 + 1 \, / \, \rho_2) = E^2 \, / \, 8\pi \qquad\qquad (1)$$

where $\rho 1$ and $\rho 2$ are the main radii of curvature. Since the only radius of curvature of the conical surface is inversely proportional to the distance R from the vertex, it follows from (1) that in equilibrium the electric field decreases proportionally to $R^{-1/2}$. Equilibrium is carried out in a field with potential

$$U = CR^{1/2} \, P_{1/2} \, (\cos \, \theta), \qquad\qquad (2)$$

where C is a constant, and $P_{1/2} (\cos \, \theta)$ is the Legendre function. From the condition of equipotentiality of the cone

$P_{1/2} (\cos \, \theta) = 0$, $\theta = \theta_o = 130.7$ ° and $\alpha = \alpha_o = 49.3$ ° are determined. Deriving from (1), the equality

$\gamma \, \text{ctg} \, \alpha_o \, / \, r = [1 \, / \, r \, (dU \, / \, d\theta)] \, 2 \, / \, 8\pi$ (3)

allows to determine the field normal to the surface of the Taylor cone -

$E = (1 \, / \, r) \, (dU \, / \, dr) = 1.4 \, \, 10^3 \, \gamma^{1/2} \, R^{1/2} \, \text{V} \, / \, \text{cm}$, (4)

a as well, calculate the critical potential difference necessary for the formation of such a cone:

$$U = 1.4 \cdot 10^3 \, \gamma^{1/2} \, R_o^{1/2} \text{ V} \qquad\qquad (5)$$

here $R_0$ is the value of R at $\theta = 0$ is the distance from the top of the cone to the extractor, the shape of which is given by the relation

$$R = R_o \, [P_{1/2}(\cos\theta)]^2.$$

A number of experiments confirm the transition from a spheroidal surface to a conical one at a certain critical field [16]. Liquid metal emitters hardened in the form of Taylor cones were actually observed [27], although in some cases [28] the angle at the apex significantly differed from $2\alpha o$. The use of an electron microscope with an electron energy of 200 keV made it possible to study the shape of the needle tip directly in the process of emission and fix a small conical protrusion with an angle of about 90 ° on the tip.

. A photograph in a transmission microscope of the formation of a Taylor cone during the occurrence of EHD emission is shown in Fig. 3: a-before the occurrence of emission, b-after the occurrence of emission.
. It is important to note that the stability of the ideal Taylor cone and the uniqueness of this solution have not yet been investigated. In experiments and photographs of working EHD emitters, configurations close to the Taylor cone as shown in Fig. 3 are clearly visible, although the shape of the electrodes is far from ideal.

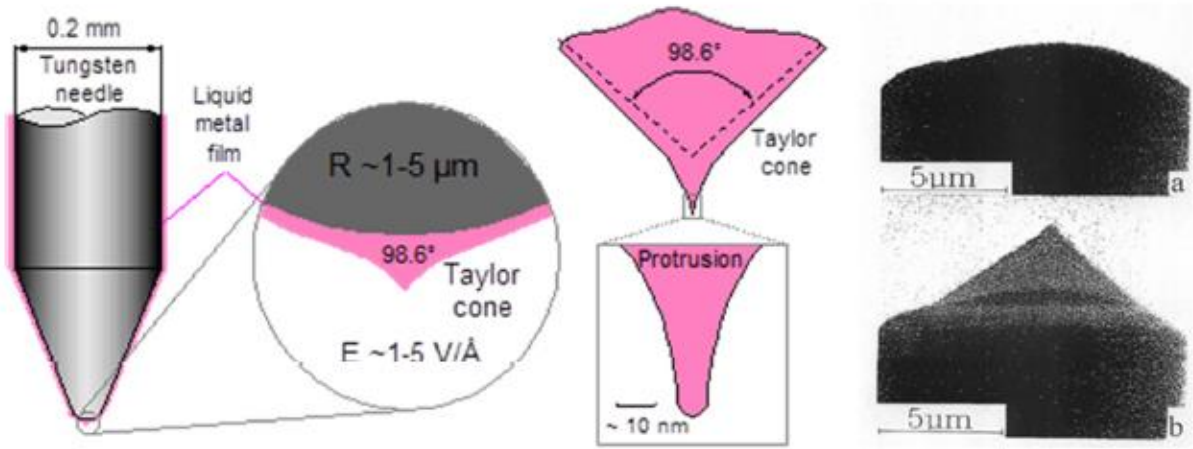

Fig. 3. The formation of the Taylor cone during the occurrence of EHD emission: a-before the occurrence of emission, b-after the occurrence of emission, photograph in a transmission electron microscope.

Now, apparently, it is generally accepted that the main mechanism for generating current in EHD emitters is the field evaporation of ions from the surface of a liquid in a strong electric field.

Field evaporation is a thermally activated process, the current density of which is described by the equation [15]:

$$j = e\sigma\omega \exp(-Q/T) \qquad (6)$$

where $\sigma \sim 2 \cdot 10^{15}$ cm$^{-2}$ is the surface density of atoms; $\omega \sim 5 \cdot 10^{12}$ c$^{-1}$-characteristic atomic vibration frequency, Q - activation energy; T -, temperature, expressed in units of energy.

A graph of potential energy describing the field evaporation mechanism for the so-called "depicting potential" model is shown in Fig. 4. The model of the depicting potential is quite well applicable to metals, and the activation energy in this model is calculated using the formula [26]

$$Q = Ho + I - e\phi + (e^3 E)^{1/2} \qquad\qquad (7)$$

where Ho is the sublimation energy of the atom, I is the ionization energy, $\phi$ is the work function, E is the evaporating field. The energy deficit of evaporated atoms can be calculated by the formula

$\Delta E = Ho + I - e\phi - Q$. (8)

For gallium, if we put Q ~ 0.1 eV, we have $\Delta E = 4.8$ eV [29], which agrees well with the value
$\Delta E = 4.6$ eV, found experimentally [30,31]

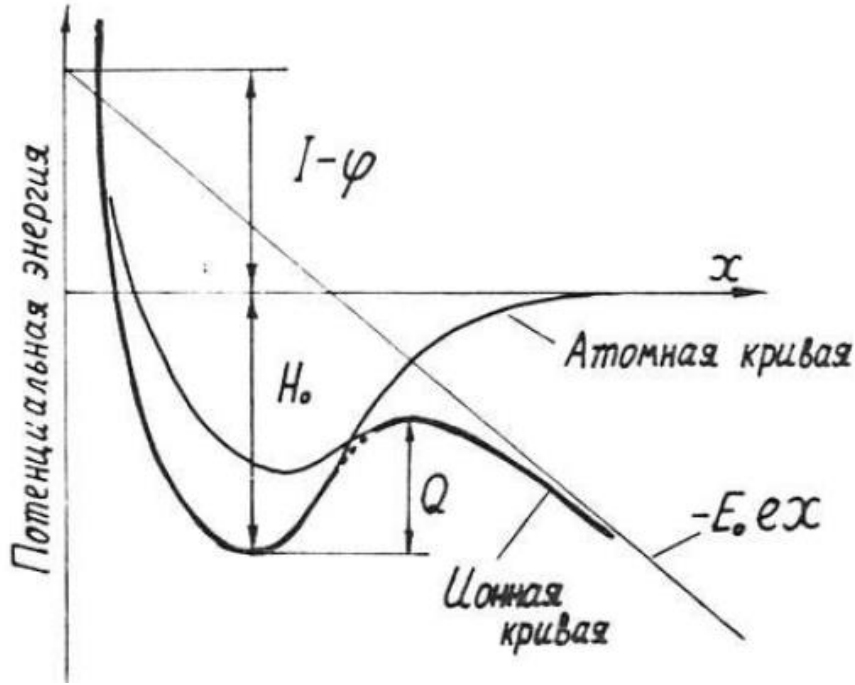

Fig. 4. Potential energy curves for the "depicting potential" model. When moving away from the surface, the atom passes into the ionic state, followed by evaporation through a barrier of an imaging potential with a height

It can be seen from (6) and (7) that the ion current density increases exponentially with increasing
The electric field on the surface of the liquid. Therefore, it is often convenient when calculating
to assume that when the current density changes in a wide
the field remains constant and, in the case of a gallium emitter, lies in the range 1.5–1.6 V / A [15].
Nevertheless, the existence of the Taylor cone is still in doubt [32,33] in a number of works, however, these works do not seem reliable [34]. In a slotted EHD emitter, a large number of Taylor cones are formed along the gap, the distance between them is approximately equal. This situation is considered in sufficient detail in [21]

LIQUID CONE FORM TAKING INTO ACCOUNT THE EMISSION OF IONS.

The main problem that arises when trying to apply the theory of field vaporized to EHD emission is the need to have a very small radius of the tip to obtain the necessary magnitude of the electric field .. In this case, the current density and space charge are extremely large, and the calculated voltage, which must be attached to EHD an emitter should be unreasonably high. However, this problem is easily removed if it is assumed that the shape of the cone in the top of the vertex is distorted, and a protrusion (stream) appears from which the emission of ions proceeds. Since the vacuum electric field at the end of the protrusion is much larger than the field at the smoothed apex: cone, it is possible to coordinate the calculated and experimentally found working voltage of the EHD emitter. The EHD model of the emitter in the form of a liquid cone with a protrusion at the apex was calculated numerically in [35]. Factors such as spatial charge distribution and fluid flow in the cone were taken into account. The calculated current – voltage characteristics and angular distributions of the emission current are in good agreement with experiment. This model (a cone with a protrusion at the top) is also confirmed in the results of [36], in which the heating of a long protrusion was calculated

~ 30 A in diameter, from the end of which there is a plei ion evaporation. The method of analytical calculation of the length and diameter of the emitting protrusion proposed in [37]

FLUID FLOW TO TAYLOR CONE

In capillary EHD emitters, fluid flows inside a relatively thick capillary; therefore, in these emitters, only a small negative pressure in the Taylor cone is sufficient to provide it with liquid. However, surprisingly, capillary emitters turn out to be significantly less stable than needle ones, because the liquid column in the capillary often breaks [38]. In needle EHD emitters, needles are commonly used that are sharpened by electrochemical etching. When etching with an alternating current, rough needles with grooves along the needles with a depth of ~ 1 μm are obtained; when etched with direct current, the needles are obtained mirror-smooth. The leakage of fluid to the Taylor cone in these two cases occurs in different ways. in rough needles, fluid flows to the tip along the grooves. The EHD emitter needles are a-smooth, b-rough with longitudinal grooves shown in Fig. 5.

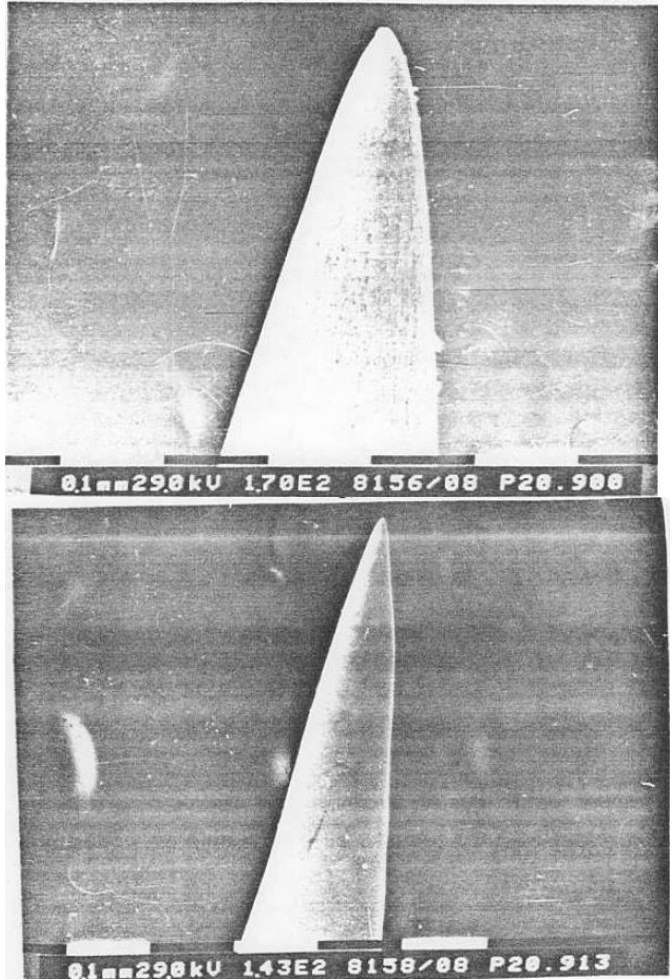

Fig. 5 . EHD emitter needles are α-smooth, β-rough with longitudinal grooves [40].

Near the reservoir, the grooves are filled with liquid, and the pressure in them is zero. Near the tip of the groove, almost empty can be, then the pressure in them is p = —y / g, where y is the coefficient of surface tension, r is the radius of the semicylindrical groove. It was shown in [11] that the grooves on a typical needle with a large margin provide fluid leakage to the Taylor cone, the maximum fluid flow corresponds to an ion current J ~ 10 mA. On smooth needles, fluid leaks due to the so-called "proppant" pressure [39]. Proppant pressure occurs in liquid films of thickness d <100 A, since the formation of such films is energetically unfavorable. Accurate calculation of the flow in this case is difficult due to the lack of data on the value of the proppant pressure for metals, but it is obvious that the hydrodynamic impedance of smooth needles should be much larger than for needles with grooves. In Fig. Figure 6 shows the current – voltage characteristics of EHD emitters with a rough (1) needle and a smooth (2) needle.

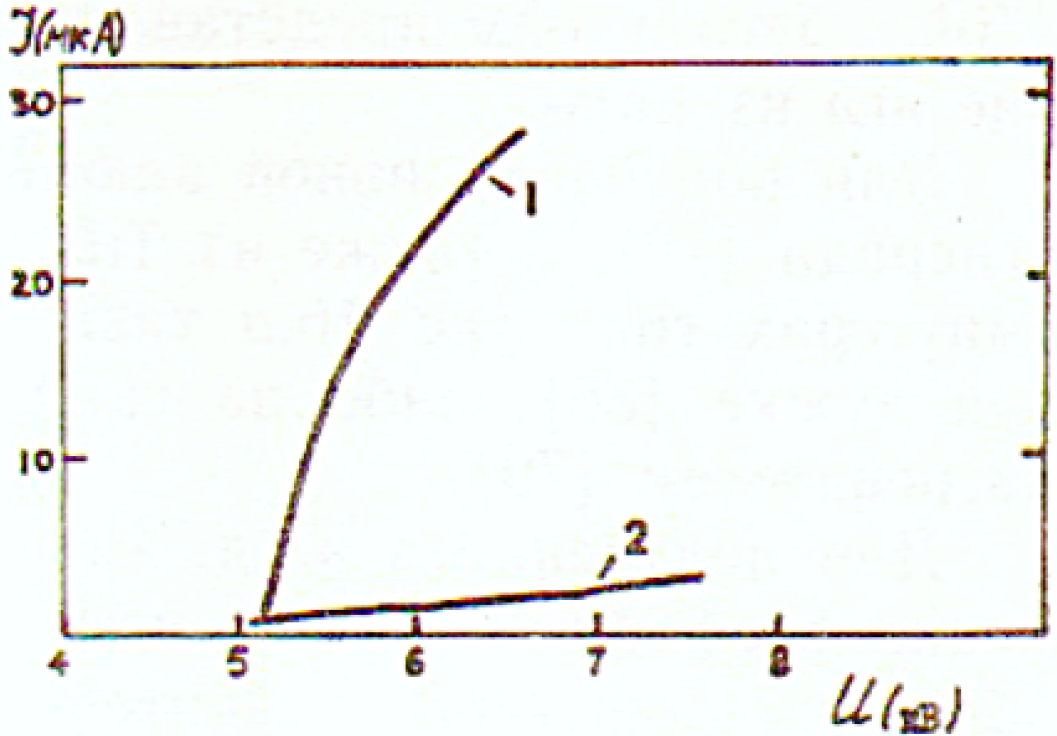

Fig. 6. Volt-ampere characteristics of EHD emitters with a rough (1) needle and a smooth (2) needle.

In many works [40,41], the slope of the current – voltage characteristics of EHD emitters is associated with the hydrodynamic impedance of the needle. It seems to us that for smooth needles this assumption is valid (see, for example, [40]), but for rough needles the impedance is so small that in all cases its influence can apparently be neglected.

VOLT-AMPER CHARACTERISTICS OF EHD EMITTERS

In Figure 7 shows a typical current – voltage characteristic of an EHD emitter of gallium ions [42]. Emission occurs abruptly at voltage U2 and disappears at voltage U1. The hysteresis at U1 <U <U2 is explained by the fact that in this range there are two stable states of liquid on the tip: in the form of a film and in the form of a Taylor cone. In [43,44] it is assumed that the inclusion of emission with increasing voltage occurs gradually. However, in experiments with capillaries and blunt needles, the authors observed the formation and disappearance of the Taylor cone visually (microscope). It was clearly seen that the Taylor cone is formed and disappears abruptly, and any no other static fluid configurations were observed. The effect of electrode geometry on the current –voltage characteristics of EHD emitters was studied in detail in [45]. In the same work, it was shown that the angular intensity of the ion beam depends not only on the emission current, but also on the focusing properties of the gap between the needle and the extraction electrode, and for EHD emitters with different electrode geometry is determined by the operating voltage of the emitter.

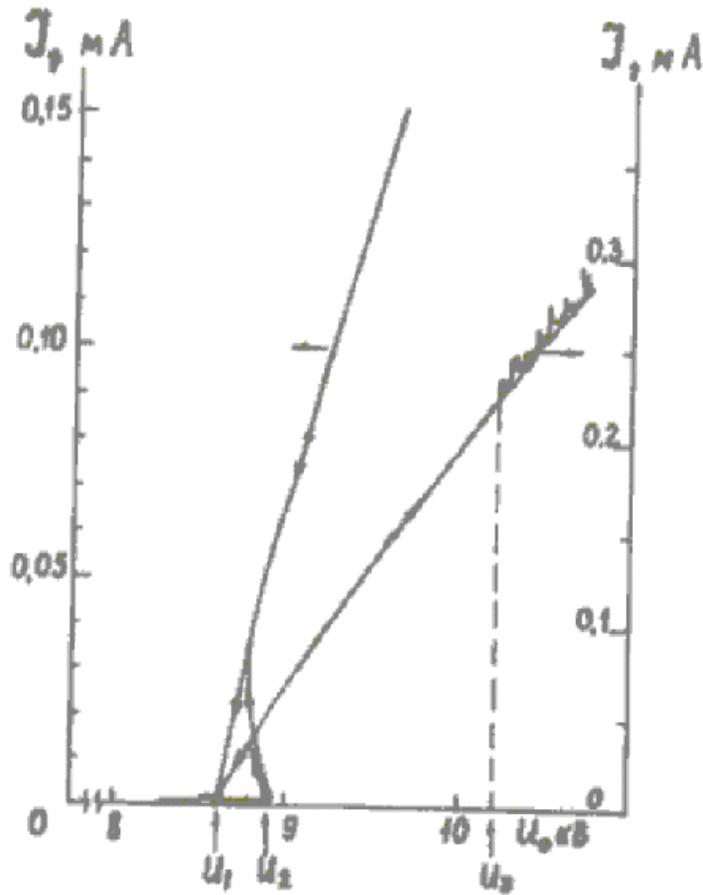

Pic. 7. Typical volt-ampere characteristic of an EHD emitter: U1, U2 - voltage off and on emission; U3 is the threshold for the development of intense low-frequency emission fluctuations.

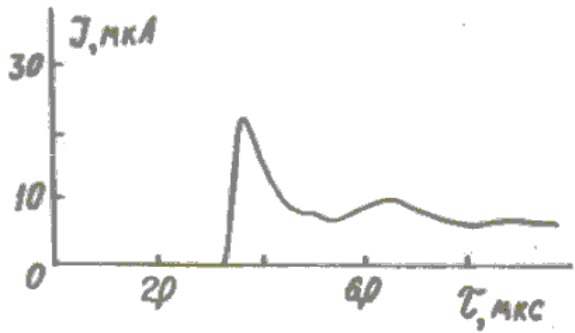

Fig. 8. The leading edge of the current pulse of the EHD emission.

At the moment the emission current appears, a current surge occurs with an amplitude J = 2Jo, where Jo is the steady-state current (Fig. 8 [46]). After several oscillations, the emission current takes a stationary value. This behavior of the ion current is apparently associated with the excitation of many modes of hydrodynamic vibrations of the Taylor cone at the time of its formation. The oscillogram of emission switching shows that the high-frequency modes decay faster and the low-frequency modes slower, as expected, based on the viscous attenuation mechanism.

# ION EMISSION, SIZE OF EMISSION ZONE

Taylor's solution is obtained for an ideal cone with an infinitely sharp vertex. In real liquids, at high field strengths, ion emission begins, and the apex of the Taylor cone is rounded with a radius of curvature r. It is now generally accepted that at low currents (J <10 µA), field evaporation of ions from the surface of the liquid predominates. It is possible that at high currents (J> 100 µA), field ionization of evaporated atoms and gas discharge in a vapor cloud near the tip can become noticeable. In [47] you suggested that at currents J> 40 µA explosive emission from the end of a liquid jet at the top of the Taylor cone can occur. . Although this assumption contradicts the results of calculating the heat balance of the jet [48], thermal explosion of the tip at high currents seems, in principle, quite possible. But the assumption of a number of authors [49] on the generation of ions even at low emission currents in a plasma cloud near the tip can hardly be accepted. During field evaporation, the ion current density can vary over a wide range, with an almost constant electric field strength on the emitting surface. Therefore, knowing the strength of the evaporating field Eo and the surface tension coefficient y, it is possible to calculate the radius of curvature of the emission surface from the Bernoulli equation. This calculation was done in [50] for an incompressible fluid. However, in [51] it is stated that tensile stresses are extremely high near the emission surface, and therefore compressibility must be taken into account. In [51], an attempt was made to calculate the energy expended in stretching a fluid through the density of surface electronic states. It seems to us that it is easier (and more correct) to calculate this energy, estimating the elasticity of a liquid by the speed of sound in it. These estimates give the value of the additional term in the Bernoulli equation with respect to the basic order (v / 2c) 2, where v is the fluid velocity, and c is the speed of sound.

. For gallium and indium, this additive in the worst case does not exceed one percent. It follows from the results of [50] that, at low currents, the radius of curvature is constant, ro ~ 10–20 A for various materials (see table), and current changes occur due to changes in the current density on the emitter surface. At high currents, when the emission density reaches its maximum value, the radius of curvature increases as $J^{1/2}$ [50]. The main drawback of the calculations [50], apparently, is that they do not take into account the possible dependence of the surface tension on the electric field strength on the surface. The results of measuring the size of the emission zone [52], at least, do not contradict the calculations.

Table 1

| Element. | T c | Eo . V / A | Ro A | J A / cm$^2$ | I * µA Calculation t | I * µA experiment |
|---|---|---|---|---|---|---|
| Al | 800 | 1.9 | 13 | 2 10$^8$ | 1.5 | 1.4 |
| Ga | 30 | 1.5 | 12.5 | 6.6 10$^7$ | 0.47 | 0.28 |
| Ga | 800 | 1.5 | 13.2 | 1.3 10$^8$ | 1,0 | 0.7 |
| In | 330 | 1.3 | fifteen | 4.3 10$^7$ | 0.44 | 0.5 |
| Cs | thirty | 0.5 | 17 | 5.9 10$^6$ | 0.08 | 0.1 |

# GLOW

Simultaneously with ion emission in EHD emitters, a characteristic glow appears in the form of a bright dot near the tip of the needle (Fig. 1). With increasing current, the luminescence intensity increases rapidly [53]. Spectroscopic studies of the EHD emitters of gallium ions showed that mainly gallium atoms excited by an ion beam shine [54]. The observed luminescence intensity corresponds to a vapor density near the tip of ~ $10^{16}$ cm$^{-3}$ at an emission current of J = 100 μA [55], or a neutral flux of ~ 1 equivalent μA. The reason for the appearance of steam with such a high density near the emission zone is still not clear. The assumption is widespread (see, for example, [10]) that metal atoms evaporate from the lateral surface of the Taylor cone and are pulled into the region of a strong field near the emission zone, it seems to us incorrect. For gallium, for example, the vapor density is extremely low, and the pulling of atoms into the region of a strong field becomes significant only at a low (nitrogen) temperature of the tip [15]. It also turned out [56] that heating a gallium EHD emitter to a temperature of 450 ° C does not increase the luminescence intensity, although the rate of gallium evaporation at this temperature noticeably increases. One of the possible explanations for the appearance of a gas cloud, in our opinion, is that the field evaporation of an ion is a nonequilibrium process, and it is possible to transfer energy to the nearest surface atom, which leads to its evaporation. Of course, such "concomitant" evaporation occurs with low probability, but it is possible that it is responsible for the appearance of a small stream of neutrals in the form of a vapor with a high density.

INSTABILITY OF EHD EMISSIONS AT LOW CURRENTS

Many studies have noted the existence of a minimum emission current J * ~ 0.1 - 1 μA, below which the emitter turns off [40,57]. As shown in [58], the emission is switched off due to the loss of dynamic stability of the liquid – vacuum interface. For its stability, it is necessary that, with a random small change (for example, an increase) in the electric field strength, the withdrawal of a substance in the form of ions increases more than its influx in the form of a liquid. Then in this place a hole is formed in which the electric field decreases to its previous value.

To fulfill this condition, it is necessary that the emission current density j be not lower than a certain threshold j *. Knowing j * and the size of the emission zone, we can estimate the minimum current J *. From the calculations [59] also follows the dependence of the minimum current on temperature, J * ~ $T^{1/2}$. The results of calculations and experiments for various substances are given in the table. And although it was reported in [60] that a stationary emission current was obtained that was much smaller than J *, this result is questioned in [61].

INSTABILITY AT HIGH CURRENT EMISSIONS

When studying the EHD emitters of gallium ions, it was found that, at an emission current above the threshold J> 30 μA, intense spontaneous emission oscillations with characteristic frequencies up to 100 MG develop [42], while emission is very stable at lower currents. In Figure 9 shows the dependence of the intensity of oscillations on the emission current [42].

A similar instability was found when working with gallium [42.59], gold [53], Li BO$_2$ melt [32], tin [62], and B-Ni-Si alloy [50]. The author also observed this instability when working on lithium, aluminum, and bismuth. It is natural to connect the development of emission oscillations with the development of capillary waves on the liquid surface of the Taylor cone. Frequencies

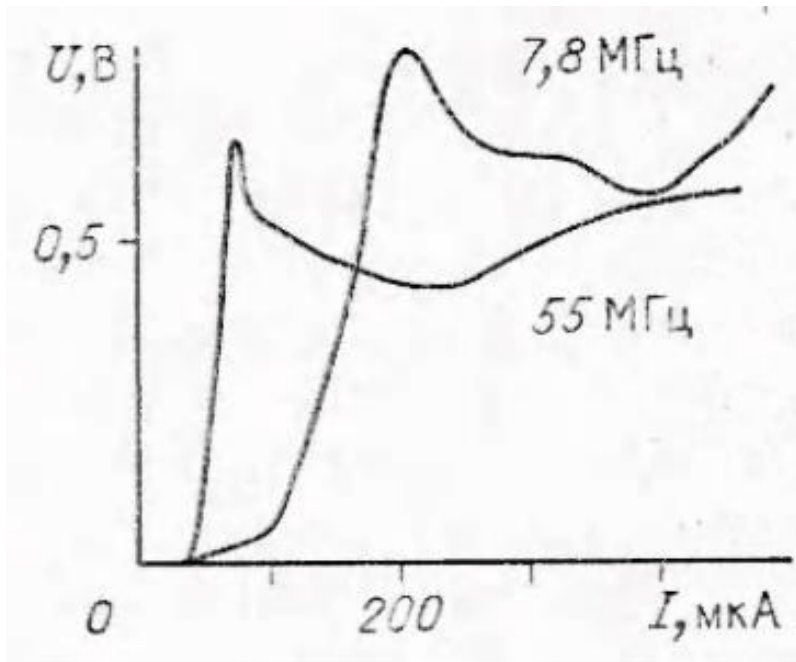

Fig. 9. Dependence of the intensity of oscillations with a frequency f = 5 5 MHz and f = 7.8 MHz (2) on the emission current [42].

f = 10-100 MHz correspond to capillary waves on the surface of liquid gallium with a scale of 1 / k ~ 0.07-0, 3 μm. The spectra of current oscillations of the EHD of a gallium ion emitter at currents of 1–20 μA, 2–80 μA, and 3–220 μA are presented in Fig. 10. The excitation of capillary waves naturally explains the generation of clusters in EHD ion emitters [11,63,64]. During intense vibrations, the top of the Taylor cone can come off from time to time, forming a cluster (clusters) of radius r ~ 1 / k (k ~ $(\rho\omega 2 / \gamma)^{1/3}$). With increasing current, lower frequency oscillations develop, and the average cluster size grows. The mechanism of excitation of these oscillations remains unclear.

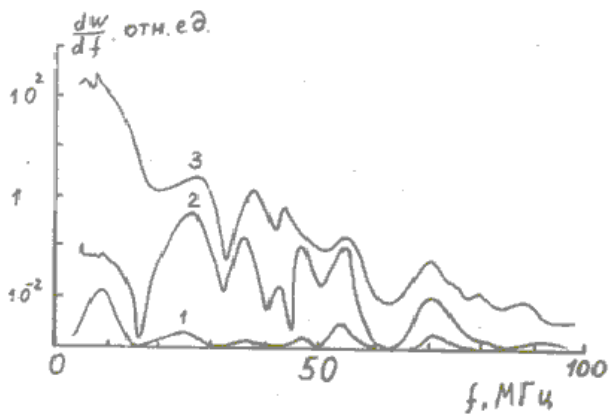

Fig. 10. Spectra of current oscillations of the EHD of the emitter of gallium ions at currents of 1-20 μA, 2-80 μA, 3-220 μA.

In [44], it is assumed that emission at high currents occurs in the "relaxation oscillations" mode, and relaxation occurs due to a thermal explosion of the emitting tip (jet). However, as indicated above, the calculations of the thermal regime of the tip presented in [45] exclude this possibility. Cluster emission and the development of oscillations were studied in [63,[65],[66],[67],[68],[69]], and the authors of [69] found an increase in the cluster output upon modulation of the emitter voltage with a frequency f = 4 kHz.

Studying: the beam collector using an electron microscope [67.68], it was possible to detect droplets ranging in size from I5 A to 1-2 μm, and it turned out that with increasing current the average droplet size increases. It was convincingly shown in [67] that the source of clusters is the Taylor cone and not the lateral surface of the needle as was stated in [10], and the clusters fly in a narrow beam (with a half-angle of ~ 2 °) in the direction of the cone axis. In the same work, it was established that the virtual (apparent) size of the cluster emission bands is less than 8 μm.

In the formation of ion beams, clusters interfere with normal operation, requiring the use of a mass filter. Recently, however, the property of EHD emitters to generate clusters has been used to deposit thin films [[70],[71],[72]]. EHD emitters with a blunt illumination (tip radius R ~ 50 μm) in high current mode (J ~ 500 μA) are used. Good adhesion of such films is noted.

TRANSITION PROCESSES

Transient processes in EHD emitters were studied in [42,53,[73], 46,[74],[75]]. With increasing voltage at the EHD emitter, emission does not occur immediately, since there is not enough liquid on the tip to form a Taylor cone. After the fluid flows to the tip after a time τ ~ 0.1 --100ms, a Taylor cone is formed and ion emission occurs. The duration of the front of switching on the emission is ~ 0.1 μs [42]. The delay in turning on the emission depends on the magnitude of the applied voltage [73,46]. The author also investigated the dependence of the intensity of the characteristic glow on time when the emitter is turned on [53].

It turns out that, due to the finite time of flight of the ions, the glow appears ~ 150 ns earlier than the current signal from the collector. The luminescence intensity front has approximately the same form as the current front; therefore, it can be concluded that the luminescence intensity monitors the emission current with a delay (or advance) of less than ~ 50 ns.

Numerical calculations of the development of instability of a flat liquid surface in a strong electric field were carried out in [[76],[77]], and in [77], taking into account the peculiarities of the development of instability, a pulsed wide-aperture EHD source of lithium ions with a pulse duration of ~ 10 ns , current density ~ 5 kA / cm2 with a total emission surface area of up to 800 cm$^2$ was proposed.. Studies of the inertia of liquid metal cathodes are described in [[78]].

KINETIC EFFECTS IN ION BEAMS

The most important parameter determining the limiting possibilities of focusing ion beams is the brightness of the source

$$B = 2 j_o W / \pi Ti$$

Here, j0 is the emission current density, Ti is the temperature of the ions on the surface of the emitter, and W is their energy after acceleration. Brightness makes sense of the magnitude of the current emitted from a unit area to a unit solid angle. In order to be able to compare different sources regardless of ion energy, the so-called normalized brightness is sometimes introduced

$$Bn = 2 j_o mc2 / \pi Ti$$

where m is the mass of the ion, s is the speed of light. If only conservative forces act in the system, then, according to the Liouville theorem, the brightness in a focused spot will be equal to the brightness of the source.

Let us evaluate the brightness of the EHD emitter of ions. As shown above, the emission density of the EHD emitter of gallium ions is $j_o \sim 10^8$ A / cm$^2$. In field evaporation, the energy dispersion of ions is determined by quantum effects upon separation of electrons and amounts to $\Delta E_{1/2} \sim 1.5$ eV [79] i.e. at the moment of emission, the ion receives a "push" in the direction of motion, and the $\Delta E$ value of this shock is random and lie in the range $-0.75 < \Delta E < 0.75$ eV. It is reasonable to assume that the ion can also receive the same impulse in the lateral direction; therefore, an estimate of $T_i \sim 1$ eV is possible for the ion temperature. Thus, we have the normalized brightness of the EHD emitter of gallium ions
$B_n \sim 10^{19}$ A / cm$^2$-rad$^2$
or brightness at an energy of W = 50 keV

$B \sim 3 \; 10^{12}$ A / cm$^2$ rad$^2$.
Aperture angles $\alpha \sim 1$ mrad are typical for the formation of submicron ion probes in ion-optical columns; therefore, it would seem right to expect the current density in a focused ion beam
$j = B \; \pi\alpha^2 = 10^7$ A / cm$^2$.
However, in real ion-optical systems, the current density in the focused beam is seven (!) orders of magnitude lower, $j \sim 1$ A / cm$^2$. Consider the reasons leading to such a significant decrease in brightness. In the formation of ion beams with a diameter d> 0.1 μm, the main reason for the decrease in the beam brightness is the chromatic aberration of electronic lenses. Of course, the lens does not increase the phase volume of the beam, but only distorts its shape, as a result of which as a result, ions with different energies appear in the center and at the edges of the focused beam. If the beam aperture angle is reduced, chromatic aberration becomes less noticeable, and the virtual size of the EHD emitter begins to play a role. The dissipative forces acting in ion beams ("friction" between the ions) lead to an increase in the longitudinal and transverse phase volumes, that is, to an increase in energy dispersion and a decrease in the brightness of the initial beam.

ENERGY ION SPREAD IN A BEAM

The dependence of the energy dispersion of ions on the emission current for a gallium EHD emitter is shown in Fig. 11 of [80].

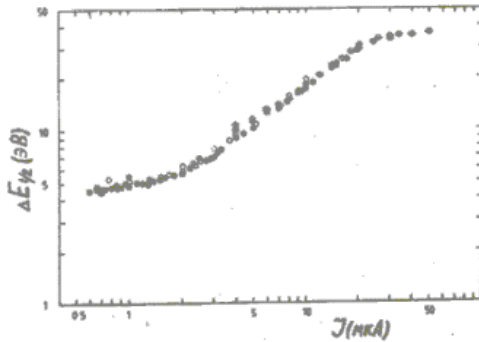

Fig. 11. The dependence of the total width of the energy distribution at half maximum on the current of the EHD emitter of gallium ions.

In field evaporation, the energy spread is determined by quantum effects and does not exceed 1.5 eV [81]. Therefore, it is reasonable to assume that the energy dispersion of ions increases due to their interaction in a dense beam, leading to heating of the longitudinal degree of freedom. It was previously believed that the heating of the longitudinal degree of freedom occurs due to the cooling of the transverse degrees during ion collisions (Bersh effect, [82,83]). Later, Nauer showed that in beams of point emitters, the transverse degrees of freedom are also very cooled, and the heating of the beam that in beams from point emitters, the transverse degrees of freedom are also very cooled, and the beam is heated due to the potential energy of interaction of ions with their random arrangement [84]. Detailed calculations of the energy dispersion were performed in [53,85], where, in particular, the dependence of the energy dispersion on the emission current for gallium was obtained

$\Delta E_{1/2} \sim J^{0.7}$

And for gold

$\Delta E_{1/2} \sim J^{0.3}$

well consistent with the experiment.

At low currents, the magnitude of energy dispersion ceases to decrease, which is consistent with the assumption that there is a minimum current density necessary for emission stability. However, the experimental dependence of the energy dispersion on the emitter temperature [86] still does not have a satisfactory explanation. Careful measurements of energy dispersion were also made in [87,88,89,90,91], and in [92] calculations of the "tails" of the energy distribution function of ions are presented.

VIRTUAL SIZE OF THE EMITTER

The virtual size of the EHD emitter was measured by the authors of [93], and the diameter at which the source brightness decreases by half with respect to the brightness in the center is taken as the virtual size of the emitter. The virtual size of the EHD emitter (a ~ 400 A) turned out to be much larger than its physical size (ao ~ 30 A), which is associated with an increase in the virtual size due to heating of the transverse degrees of freedom of ions during beam drift in the ion-optical column.

Numerical calculations of an increase in the virtual size were performed in [94], analytical in [84], and the results of analytical calculations differ from [94] and are much better in agreement with experiment. The virtual size of the emitter is important in the formation of submicron ion beams. However, the profile of the focused beam on the target is also of great importance. Measurements by various authors [95,96,97,98,99,100], show that the distribution of the current density in the beam remains Gaussian up to intensities 2-3 orders of magnitude lower than the maximum, but in the future the intensity decreases with a radius much slower, according to the law $r^{-m}$, where m = 2.5 [91] or m = 3.3 [92].

This effect is explained either by the mutual repulsion of the ions in the beam (that is, by the "tails" function H ($\beta x$) [94]), or by the scattering of ions by the residual gas [89] and, finally, by the chromatic aberrations of cylindrical lenses for particles from the "tails" »Energy distribution functions [95]. In [100], it is assumed that, with decreasing intensity, chromatic aberration begins to play a role, and only at very low beam intensities does scattering by the residual gas dominate. .

EHD EMISSION FROM DIELECTRIC MELTS

Usually, high conductivity melts of metals and alloys are used as working substances for EHD emitters. This circumstance is due the name of these sources is liquid metal ion sources. And although in [101].

The design of the source with an EHD emitter was used to obtain ions from the dielectric compounds CsCl and $P_2O_5$, the ions were generated not in the EHD mode of the emitter, but in the surface-ionization mode at the tip and in the electron-impact ionization mode. However, there are no fundamental ones: there are reasons to limit the choice of working substances of EHD emitters only to metals and their alloys. The requirements for working substances — sufficiently high conductivity in the liquid state and not too high vapor pressure — can also be met by salt melts.

We were able to obtain stable EHD emission from LiB $0_2$, NaOH, and $NaBO_2$ melts. [22]. We used a source with an EHD emitter and an ion-optical column to focus the handle, analyze its size and mass composition (Fig. 12).

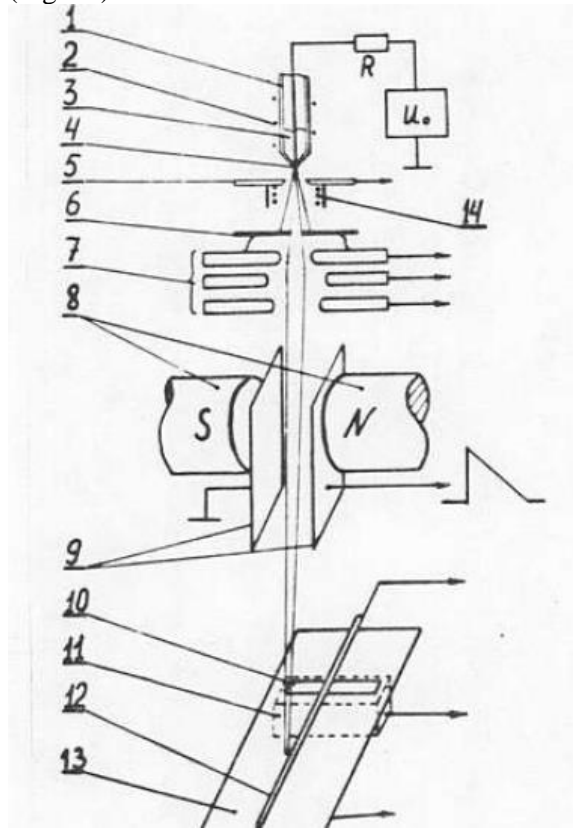

Fig. 12. Scheme of an EHD source with a focus and beam analysis system. I - a container with a working substance, 2 - a heater, 3 - a working substance, 4 - an emitter needle, 5 - a drawing electrode; 6 - an aperture diaphragm, 7 - a single lens, 8 - poles of an electromagnet, 9 - from deflecting plates, 10 - more mass spectrometer, 11 - collector, 12 - wire, IZ - screen, I4 - extractor electrode heater.

The tip of the needle 4 with a tip radius of 3-5 μm extends from the cylindrical container 1 with the working substance 3 through an opening in its conical bottom to a drawing electrode 5 with an opening for the beam. The container is heated to the melting temperature of the working substance by a heater 2 made of tantalum wire with ceramic insulation. A pulling voltage U is applied between: the needle 4 and the pulling electrode 5. A part of the beam is cut out by the aperture diaphragm 6, focused by a three-electrode electrostatic lens 7, passed through the gap of the electromagnet 8 and registered by the collector 11, which is located behind the screen 13 with a narrow slot 10 parallel to magnetic field of the mass analyzer.

Along the magnetic field, the beam can be deflected by the electrostatic deflector 9 and cross the wire 12, the signal from which serves to determine the size of the focused beam. This whole system was tested and calibrated when working with molten metals and alloys.

In these experiments, ion beams in the EGD mode of the emitter were obtained from melts of lithium boric acid ($LiBO_2$), sodium boric acid ($NaBO_2$.), Sodium hydroxide (NaOH).

The emitter needles were wetted with melts of these substances in air and pieces of the substance were loaded into the container.

However, it was possible to obtain stable EHD emission in only a few seconds, then breakdowns began in the source. It turned out that the vapors and droplets of the working substance condense on a cold drawing electrode, covering it with a thick non-conductive layer. This layer is charged by ions and breaks through, disrupting the source.

The problem was solved by applying a pulling electrode with an additional tungsten heater 14, located behind the hole for the beam. This heater warms up until a small thermionic emission heats the pulling electrode, and thermoelectrons accelerated by the voltage applied to the EHD emitter provide additional local heating of the needle. The working substance falling on the heated drawing electrode spreads and spreads along it with a thin conductive film. Due to this, it was possible to ensure long-term stable operation of the EHD emitters on dielectric melts.

Examples of the dependences of the ion current Jo to the electrode 6 and the current of the focused beam Jc to the collector 11 from the pulling voltage, recorded by a two-coordinate recorder, are shown in Fig. 13. From the characteristics of EHD emitters with metal melts obtained under similar conditions, these dependences differ by a significantly higher ion thermal emission current, which increases almost linearly with an increase in the drawing voltage U. At $U = U_2$, the melt film self-sharpens and the current Jo sharply increases due to the transition to the EHD mode of the emitter. An increase in current with a slow increase in U occurs during 100 ns. The voltage at the emitter decreases due to the voltage drop across the limiting resistance. In the region $U > U_3$, low-frequency current oscillations Jo.

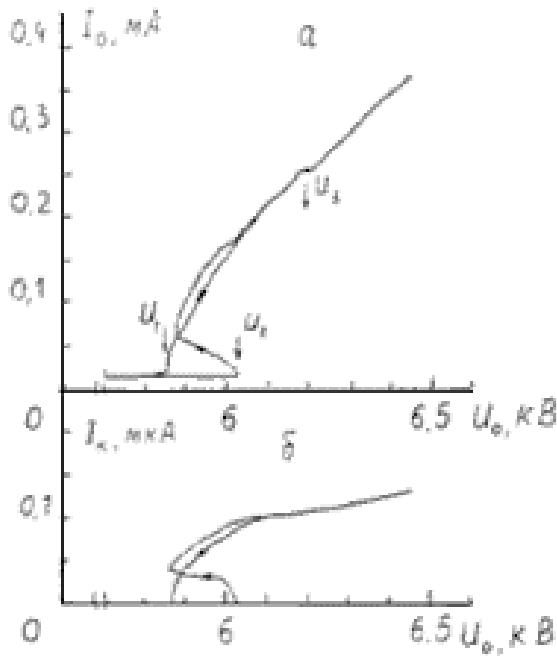

Fig. 13. Dependence of the ion emulsion current Io, (a) and the current of the focused bunch Ic (6) on the pulling voltage Uc

With a change in U in the region $U > U_1$, the Jo changes are reversible, but, in contrast to the EHD of the emitters with metal melts, there is a noticeable hysteresis. With decreasing $U < U_1$, the EHD emission mode disappears and reappears only at $U > U_2$. With a further increase in the emitter temperature above the melting temperature of the working substance, the thermionic emission current rapidly increases from $10^{-5}$ to $10^{-3}$ A. However, a clear transition to the EHD mode of the emitter is not observed, although the steepness of the dependence of J on $U_c$ in the region $U \sim U_2$ increases by an order of magnitude.

The current of the focused beam Jc to collector 11 (Fig. 13) in the thermal emissivity mode (at U <U₁) was less than the detected level (~ 10⁻⁹ A) due to the low brightness of the beam. The dependence of Jc on U is similar to the dependence of Jo on U and shows that the brightness of the beam increases significantly more than the current Jo when switching from thermal emission to the EHD mode of the emitter, but with a further increase in U, the brightness changes slightly. The resulting beams were focused on the collector (28 cm from the emitter) to a diameter of 0.03 mm and were analyzed by masses. Mass composition of beams obtained in the EHD mode of an emitter from Li B O₂ melts. and Na OH is shown in Fig. 14.

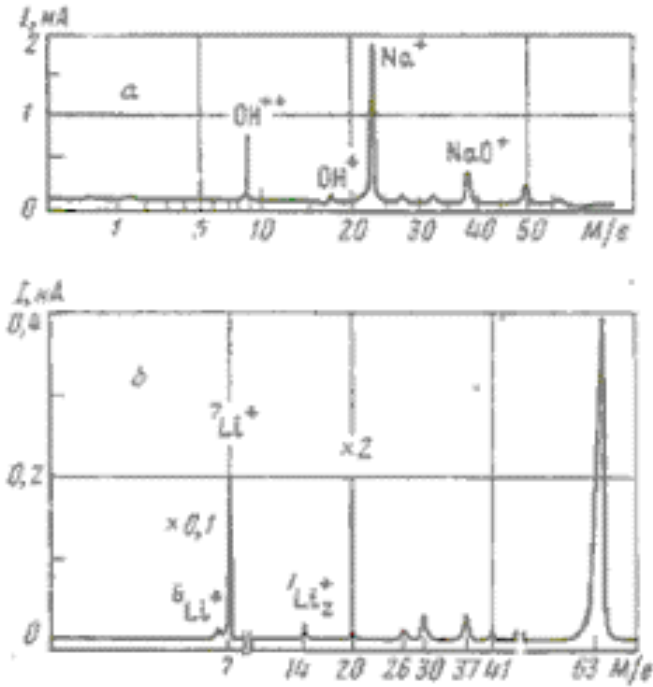

Fig. 14. The mass composition of ion beams when used as a working substance: a-NaOH b-LiBO2 (H2LiBO3) from [53].

The preparation of negative Cl-, J-, F- ions from CsCl, CsI, CsF melts is described in [25]. From the mass spectra [22] it is seen that most of all in the beam of metal ions. When working on NaOH, doubly charged OH + ions are also formed. However, nowhere in the literature did we manage to find information about the existence of such ions. Therefore, now we believe that the first nickname is probably O ++ ions, and the calibration of the mass spectrometer is not quite accurate. When using Li BO₂, a large peak with a mass of ~ 63 is formed. We associate this peak with the (BO₂ + H₂O) + ion, since LiB0₂. very hygroscopic, and him; chemical form, is often written as H₂LiBO₃. And finally, in both cases, we can conclude that during field evaporation a very "soft" ionization of the substance occurs, during which complex ions do not break into fragments and there are hopes for using field evaporation to identify complex (for example bioorganic) molecules.

PRACTICAL ASPECTS OF WORK WITH EHD EMITTERS, MANUFACTURE OF EMITTERS

We have tested various designs of EHD emitters, including EHD emitters with a capillary, "needle" emitters with arches and container-type emitters. In the end, it was decided to develop the design of a universal container-type source suitable for producing a wide range of ions, providing for the possibility of its installation in an ion-optical column.

As a rule, in our work, we used EHD emitters of the container type, the device of which is shown in Fig. 15 [53,[102]]. Container EHD emitters have significant advantages over emitters with a bow (Fig. 2) since you can lay a significantly larger supply of working substance, and the evaporation of its volatile components in container emitters is difficult. Container emitters are suitable for working with almost all working substances, however, for each substance, the selection of needle material and a replaceable container is necessary. The most versatile is a replaceable graphite container, since graphite is very poorly wetted by metal melts, although a stainless steel container is required when working with lithium and cesium.

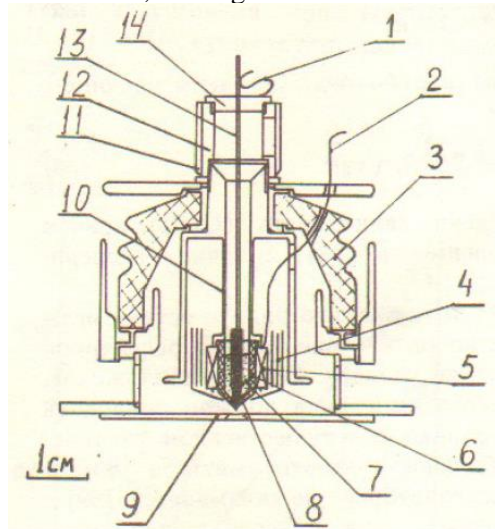

Fig. 15. Scheme of ionic EHD source: 1 - needle clamp; 2 - heater current supply; 3 — ceramic insulator; 4 — thermal screens; 5 — heater; 6 — replaceable container; 7 — working substance; 8 - emitter needle; 9 - traction electrode; 10 - case + emitter; 11 - a nut; 12 - cork; 13 - needle holder.

. The simplest heater option for a container is a 0.3 mm diameter tantalum wire insulated from pieces of a ceramic tube tightly laid on the surface of the container. However, at high temperatures, ceramics undergo strong gas separation, and the surface of the emitter needle may become contaminated. Therefore, a heater in the form of a tungsten spiral, inside which a container is placed with a small gap, is more preferable. Since there are no insulators in the heater zone in this design, the heater is cleaner, but this heater design is much more complicated to manufacture and less economical. Source depicted in Fig. 15, allows to obtain EHD emission when working on substances with a melting point up to 900 ° and with a relatively high vapor pressure. In this source, when selecting the appropriate needle and container materials, aluminum, boron, phosphorus, antimony, lithium, nickel, copper, gallium, indium, cesium, gold, lead, bismuth, silicon, tin ions were obtained. The source needles were made of tungsten, nichrome, nickel, glassy carbon and lanthanum hexaboride. Metal needles are made by etching from the corresponding wire with a diameter of 0.3-1.5 mm. The end of the wire is lowered into the bath with electrolyte to a depth of 5-10 mm and etched with alternating current at a voltage of 5-10 V to obtain a tip. The process lasts about 10 minutes.

Lanthanum hexaboride needles are pre-ground on a diamond wheel and then etched in the same way as metal needles. Glass carbon needles are also pre-ground on a diamond wheel, and then brought into the flame of an oxygen-hydrogen burner. To obtain an oxygen-hydrogen mixture, the author used an electrolyzer with a capacity of about 100 watts. As a rule, when preparing a new source, the emitter needle was tin-plated in advance in a superheated melt of the working substance in vacuum.

For tinning, the needle is mounted on the manipulator in the vacuum chamber, after pumping out, the working substance in the crucible is heated to the required temperature (temperature is controlled using a pyrometer through the window of the vacuum chamber), and the needle is dipped into it. If the working substance at the melting temperature does not wet the needle well and does not allow significant overheating (evaporates), you can carry out tinning of the needle with the help of another EHD-emitter. The needle, which needs to be irradiated, is first cleaned by ion bombardment and doped with a beam of the desired ions, and then in the cluster generation mode it is covered with a film of the working substance.

. The served needle is installed in the container of the ion source, and pieces of the working substance are laid there. The pull electrode is set symmetrically with respect to the needle. The source is installed in a vacuum chamber so that the needle and the pulling electrode of the EHD emitter during operation can be observed using a magnifying glass through a window in the chamber. After evacuation, the working substance is brought to melting, and EHD emission is achieved by increasing the voltage. Usually, upon first switching on, EHD emission occurs at a voltage U exceeding the voltage $U_2$ (Fig. 6) by 2–4 kV. At the next switching on, the current – voltage characteristic of the source corresponds to Fig. 6 and practical does not change over time.

The emission current and the symmetry of the ion beam are controlled by a collector with a luminescent screen. When preparing the needle, its shape may deviate axially symmetric, as a result, the axis of the Taylor cone (axis of the ion beam) will not coincide with the axis of the needle. For metal needles, the deviations are usually small (<10 °), however, for needles made of non-metallic materials, the deviation of the beam axis can reach 40 - 50 ° (for particularly unsuccessful specimens). Since in this case the beam will not enter the input aperture of the ion-optical column, the emitter needle must be changed.

At high emission currents, the formation of two or more Taylor cones is possible, while a corresponding number of luminous circles appear on the luminescent screen, which may partially overlap. If an EHD source exhibits steady, stable operation and emission symmetry, it is suitable for installation in an ion-optical column.

OBTAINING IONS OF DOPING ELEMENTS

One of the applications of EHD emitters is the formation of ion beams for maskless implantation of impurities in semiconductor structures. A wide range of ions is used for these purposes, but the main alloying elements are boron, phosphorus, aluminum, antimony. Improving the sources of these elements pays great attention. For normal operation of the source, the working substance with the ionizable element must be in the molten state, moisten the needle well, and have a sufficiently low vapor pressure.

For these reasons, difficulties arise in obtaining ions of alloying elements. Boron compounds are too refractory, phosphorus and antimony are sublime before melting, aluminum melts actively interact with almost all materials and quickly dissolve source details. In more detail, the effects that make it difficult to obtain ions of alloying elements are considered in [103,104]

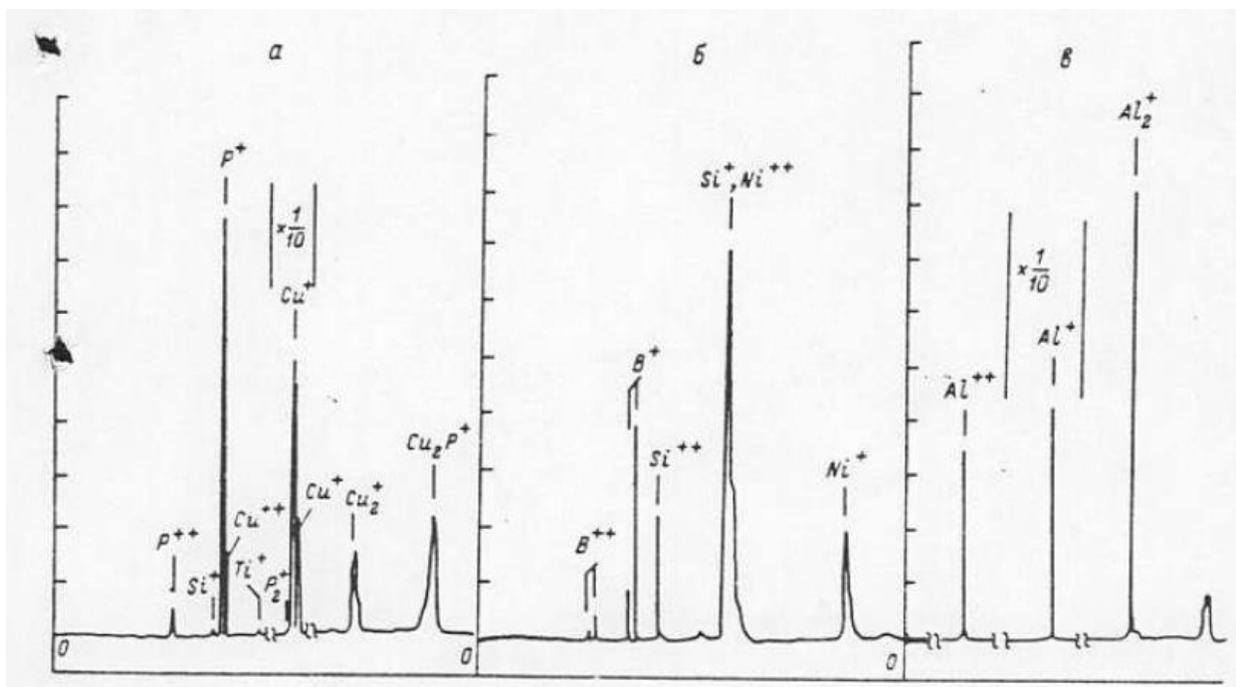

Fig. 16. Mass spectra of ion beams of EHD emitters with various working substances a-copper-phosphorus alloy, b-alloy $Ni_{45}B_{45}Si_{10}$, b-aluminum [53].

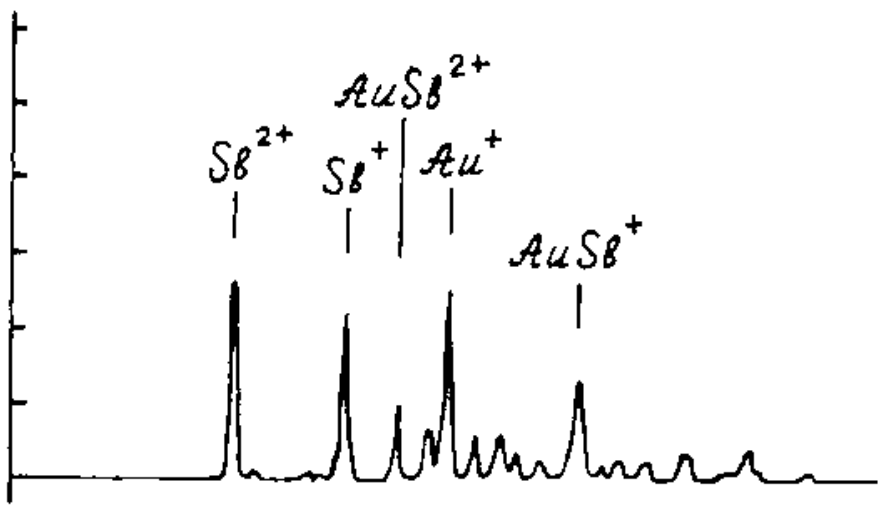

Fig. 17. Mass spectrum of the ion beam of the EHD emitter with $AuSb_2$ alloy [53]. To obtain ions of aluminum, antimony, phosphorus and boron, we used EHD sources, the design of which is described in the previous section. The source was mounted in an ion-optical column to form a beam and register the mass distribution of ions, similar to that shown in Fig. 12. The temperature of the container of high-temperature sources was measured with a pyrometer through the window of the vacuum chamber. Heaters of low temperature (T <800 ° C) sources were pre-calibrated using thermocouples.

To obtain phosphorus ions, a MP copper-phosphorus alloy with a phosphorus content of up to 9% by weight and a melting point of ~ 800 ° C was used. Tungsten needles with radii of curvature of ~ 5 μm were dipped into the MP melt in air and installed in a source container with an additional supply of working substance. Using the container allows you to significantly increase the supply of working substance and prevents the evaporation of volatile impurities from the melt. Of great importance is, its choice of container material.

In stainless steel containers, copper-phosphorus alloy leaked up the walls due to good wetting of the container, which limited the life of the source to 8 hours. The quartz containers were not wetted by the melt, but charging quartz with secondary electrons from the extracting electrode caused breakdowns on the surface and disrupted the stable operation of the source. The best results were obtained with graphite containers. The current-voltage characteristics of the emitter are similar to those shown in Fig. 6. An example of the mass of the beam spectrum is shown in Fig. 16 .

At an emission current of up to 30 μA, the current of P + ions was 15% of the current of Cu +. With a further increase in the emission current, the fraction of P + ions decreased, but was restored after a decrease in current. For 40 hours of operation of the source with multiple melting and solidification of the working substance, the current-voltage characteristics and the mass composition of the beams remained practically unchanged.

To obtain aluminum ions, needles from lanthanum hexaboride and glassy carbon were used, which were installed in graphite containers.

The needles from lanthanum hexaboride were sharpened mechanically to a tip radius of 20 μm, then brought by electrochemical etching in dilute hydrochloric acid at an alternating voltage of about 8 V. These needles are well wetted by aluminum, but also dissolve in it, which limits the duration of the source until the needle is replaced at 10 hours. Glass carbon needles were sharpened to a tip radius of 10 μm mechanically and brought to a radius of 5 μm in an oxygen-hydrogen flame. Glassy carbon needles are significantly more resistant to molten aluminum, but there is evidence in the literature about the difficulty of wetting such needles. In [36,37] for Nickel or titanium was preliminarily sprayed onto the needle by applying an aluminum film; however, the emitter did not succeed in prolonged operation (more than one hour). We managed to ensure good wetting by lowering the needle into the aluminum melt, heated in vacuum to 1000 ° C. At this temperature, there is a weak mutual dissolution of carbon and aluminum. and cohesive wetting of the needle is provided.

The service needle was installed in a graphite container with an additional supply of aluminum. The source worked at a container temperature of 700 ° C. The mass composition of the beam is shown in Fig. 16. For 40 hours of operation of the source with an emission current of 40 μA, its characteristics have not changed.

To obtain boron ions, we used the $Ni_{45}B_{45}Si_{10}$ alloy with a melting point of 900 ° C, proposed in [105]. The emitter needles were made of glassy carbon and lanthanum hexaboride. To achieve good wetting, glass-carbon needles were pre-tinned with aluminum. The needles made of lanthanum hexaboride were well wetted by $Ni_{45}B_{45}Si_{10}$ alloy at high temperature (T> 1100 ° C).

The EHD emitter of boron ions with a glass-carbon needle worked normally until the first shutdown (in our conditions, no more than 8 hours). However, after the first cooling-heating cycle, its characteristics changed, and after the second cycle, it usually stopped working at all the characteristics changed, and after the second cycle, he usually stopped working at all. Under the microscope, it was seen that the tip of the needle was chipped, and the alloy was lagging behind flakes from glassy carbon. Apparently, this is due to the fact that $Ni_{45}B_{45}Si_{10}$ and glassy carbon have different coefficients of thermal expansion, and since both of these substances are brittle and are not capable of plastic deformation, the resulting mechanical stresses and lead to delamination of the alloy. The data given in reference [103] on the operation of a source of boron ions with a glassy carbon needle for 250 hours, apparently, relate to continuous operation without turning off the source heater.

EHD emitters of boron ions with lanthanum hexaboride needles allowed multiple cooling-heating cycles and worked stably for a long time. A typical mass spectrum of the ion beam is shown in Fig. 16. Such an emitter was tested for 24 hours with multiple on and off. During this time, the threshold voltage for the occurrence of emission decreased from 10.3 kV to 8.6 kV. Other changes in characteristics did not occur.

To obtain antimony ions, an AuSb2 alloy with a melting point of 460 ° C, proposed in Ref. [103], was used.

This alloy gives an increased yield of antimony ions, but also has a higher vapor pressure at the melting point, in contrast to the previously proposed $Sb_{50}Pb_{42}Au_8$ alloy [47] with a melting point of 300 ° C and a fraction of antimony ions in the beam of about 7%. Due to the high vapor pressure of antimony, the AuSb2 alloy was used by the authors of [106] in an EHD emitter with a porous needle. Working with the AuSb2 alloy, we used an EHD emitter with a conventional tungsten needle (Fig. 15). The mass spectrum of the ion beam, taken about an hour after the start of the emitter, is shown in Fig. 17. A large number of doubly charged antimony ions make it possible to obtain beams with an energy that is twice the operating voltage of the ion-optical column, which is often very convenient. Despite the high vapor pressure of antimony, the emitter worked stably, which is associated with the open surface of the melt.

Due to the high cost of the alloy, a very small piece of it was loaded into the container, which was only enough for 38 hours of operation. 1 hour before the supply of the working substance was exhausted, the fraction of antimony ions in the beam fell by 30% from the initial level. By laying in the container more working substance, it is possible to achieve a significantly longer duration of the source and a constant composition of the beam.

FORMATION AND APPLICATION OF SUBMICRON ION BEAMS

A simple system for microprocessing with a focused ion beam is described in [107]. Similar systems were created by a group from Ryazan [108,109,110] and in [111].

A fairly advanced system for microprocessing with a focused ion beam was created by a team from Novosibirsk [112,113].

In Fig. 18 of [53,113] shows a diagram of an ion-optical column for producing submicron ion beams. The general scheme of the installation for microprocessing with a focused ion beam is shown in Fig. 19 [53,105]. A photograph of an ion-optical column for microprocessing with a focused ion beam is shown in Fig. 20. Photograph of a focused ion beam microprocessing plant. shown in Fig. 21.

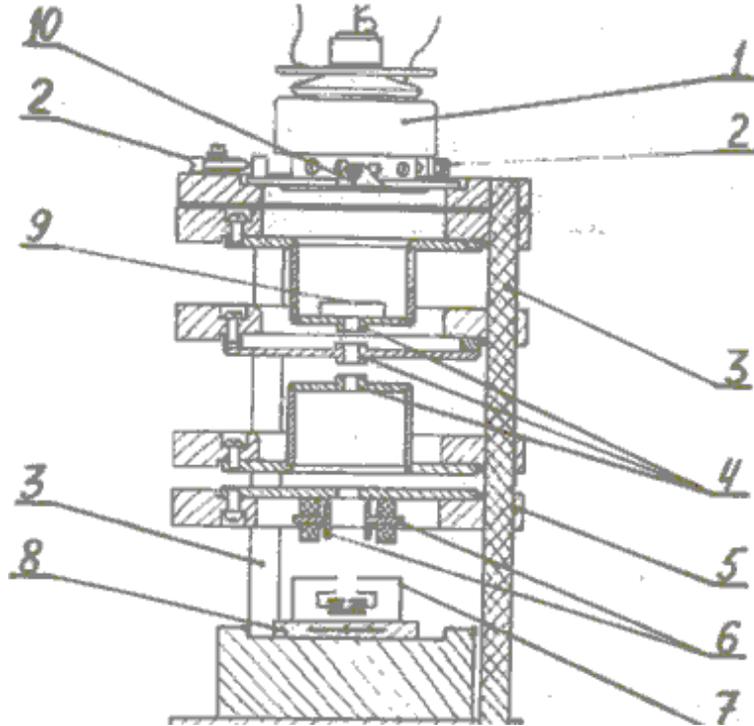

Fig. 18. Scheme of the ion-optical column [44]: 1 — ion source; 2 — source movement mechanism; 3 — ceramic rods; 4 - electrodes of a single lens; 5 — flanges; 6 — deflecting rods; 7 — secondary electron detector; 8 - movable table; 9 - aperture diaphragm; 10 - needle ion source.

The ion beam from the EHD source is limited by the aperture diaphragm 9 and focuses on the target 8. Since absolutely accurate assembly of the ion-optical column is impossible, a device is needed to compensate for the astigmatism of beam 6 (stigmatator).

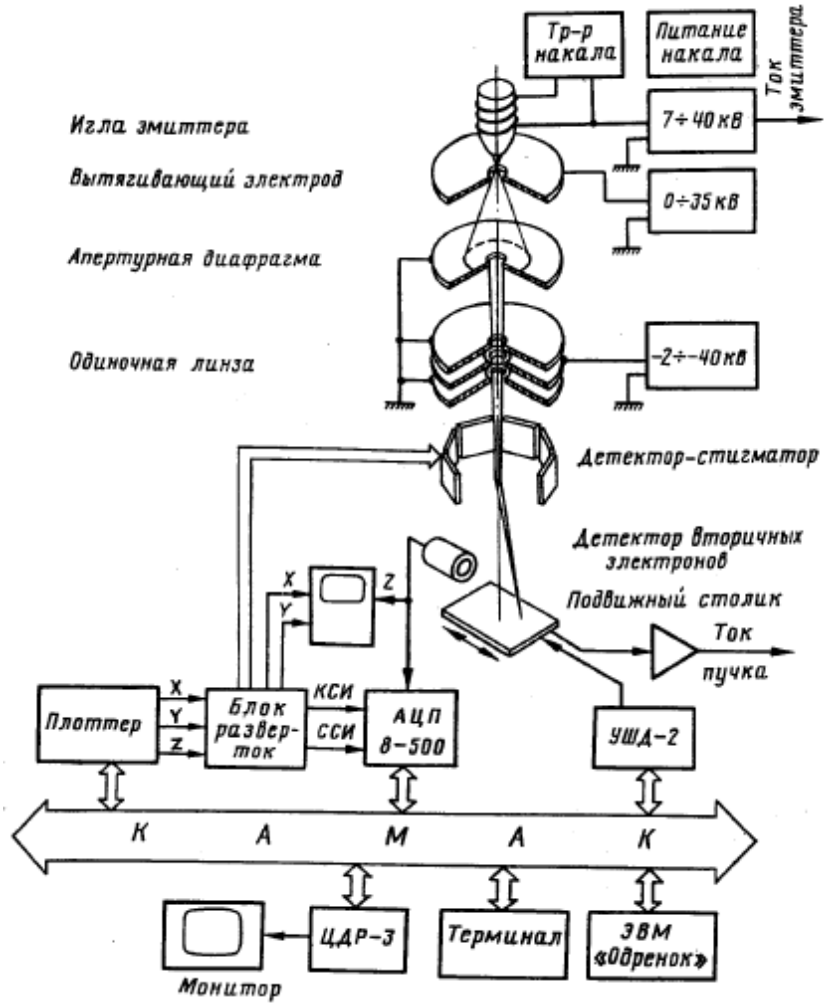

Fig. 19. The complete scheme of the installation for microprocessing by an ion beam [53].

The secondary electron detector is used when operating in the mode of a scanning ion microscope. The design of an octupole stigmator- deflector and a secondary electron detector are described in [53].

For scanning submicron ion beams, octupole deflectors are usually used, which have significantly lower aberrations than a system of two pairs of deflecting plates. Another advantage of an octupole deflector is that it can at the same time to work as a stigmator that uses beam astigmatism. The potential distribution on the electrodes of such a stigma deflector is shown in Fig. 22.

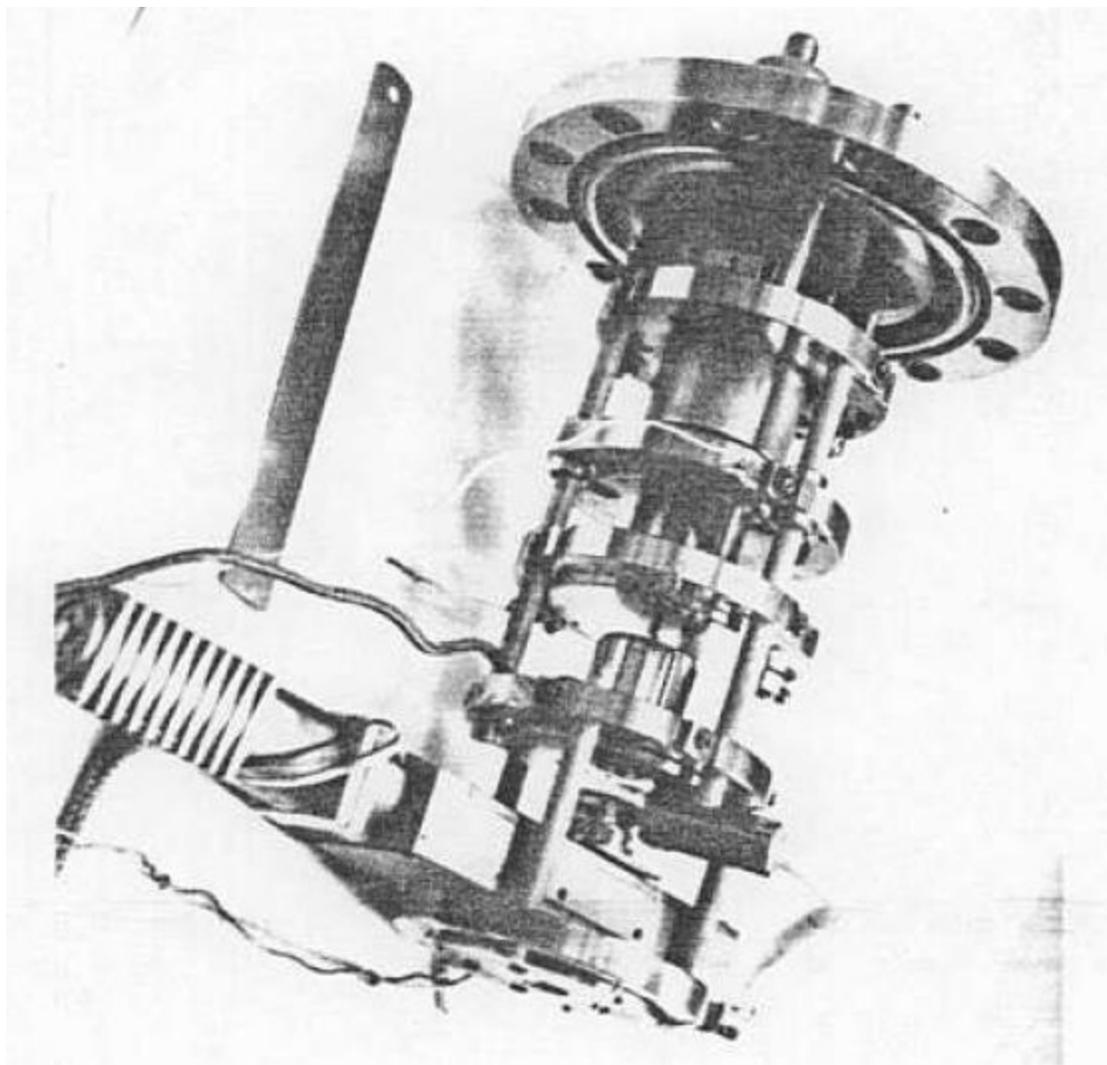

Fig. 20 . Photo of an ion-optical column. A vacuum flange is visible above, a movable table below [48].

The secondary electron detector is shown in Fig. 23. A layer of a phosphor (polystyrene with the addition of parathelphenyl) is applied to the end face 3 of the flexible fiber 4, on top of which a thin layer of aluminum is sprayed. A voltage of + 10 kV is applied to the aluminum coating. Secondary electrons knocked out by the ion beam are collected by a grid 2, having a potential of +300 V. After passing through the grid, the electrons are accelerated, and breaking through a layer of aluminum, they cause the phosphor to glow. Light is recorded using an FEU-130 photomultiplier at the other end of the fiber.

The entire design of the ion-optical column is made rigid enough to reduce vibrations, since when working with beams with a diameter of less than 1 μm, even insignificant relative displacements of the emitter needle, lens, and target deflect the beam by distances significantly exceeding its diameter. As shown in [9,53,[114]], the current density in a focused ion beam is determined by the ion energy and the chromatic aberration coefficient of the lens, and the beam diameter is proportional to the diameter of the aperture diaphragm. Under optimized conditions, ion beams with a diameter of 0.1-0.2 microns and with a current density of 1 A / cm2 at an energy of W ~ 30-100 ke [115,116,117] are obtained, although in [118] it was reported that a beam with a diameter of 0.04 microns was received, and in [119] the current density $j = 6$ A / cm$^2$ was reached.

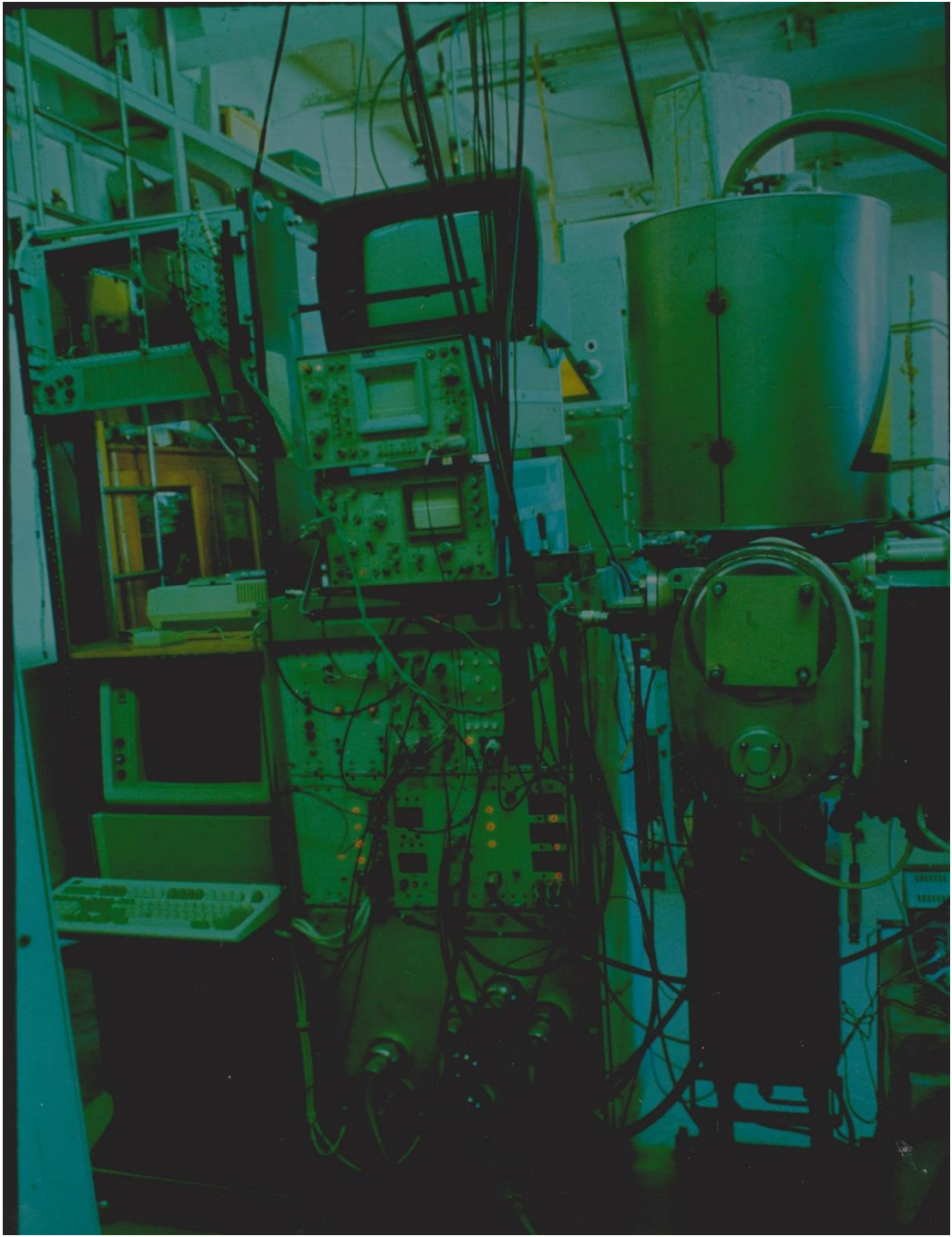

Fig. 21. Photo of the installation for microprocessing by focused ion beam at the INP, USSR Academy of Sciences. On the left is the control stand, on the right is the vacuum chamber, in the center is the power stand [53].

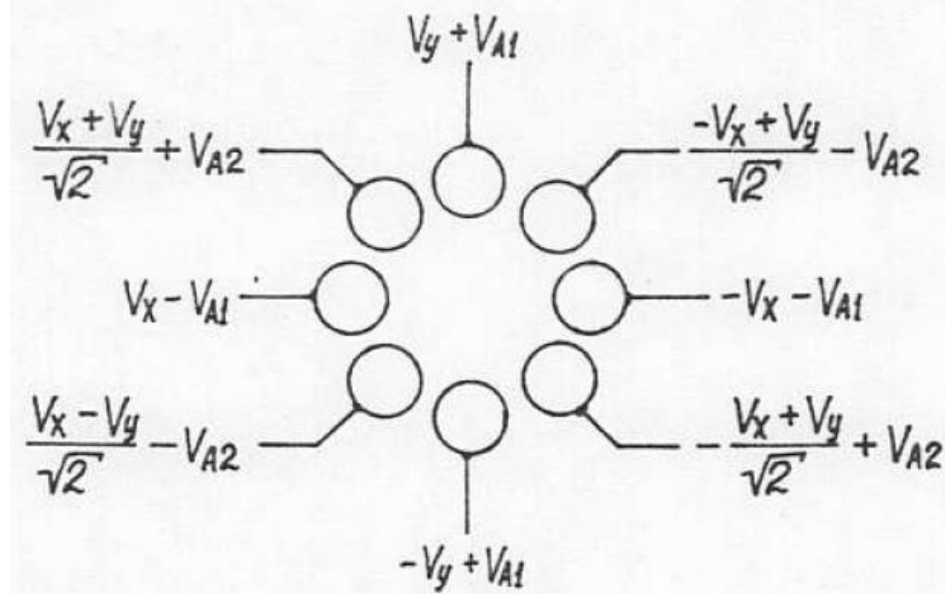

Fig. 22. Voltage distribution on the electrodes of the octupole stigma-deflector. Vx, Vy-deflecting stresses, $V_{A1}$, $V_{A2}$-stresses adjust astigmatism.

It is proposed that submicron ion beams be used for scanning electron microscopy and surface microanalysis, ion lithography, direct maskless implantation, for correcting defects in optical and X-ray patterns, and microprocessing of the surface.

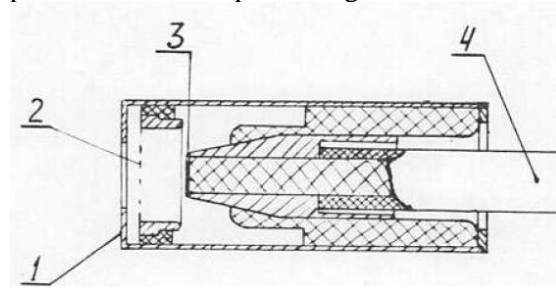

Fig. 23. Secondary electron detector. 1-casing, 2-collecting grid, 3-phosphor with a aluminum coating, "-flexible fiber.

Now the main application of installations with submicron ion beams is the correction of template defects and modification of microcircuits [120,121]. Moreover, in recent years, the method of selective ion-stimulated deposition of metals from the gas phase has been widely used to create conducting paths and eliminate light defects in templates [12]. As a material for deposition induced by an electron or ion beam, gas-chemical sources are used — heated crucibles containing low-melting organometallic precursors (W, Pt deposition) or $C_{10}H_8$ naphthalene (carbon deposition). Precursor substance in the gaseous state is transported to the sample through nozzles and adsorbed on its surface. The decomposition of the organic precursor occurs when the surface of the sample is scanned by an electron / ion beam. Scanning takes place according to a pattern, which allows you to create patterns of a given shape on the surface. In Fig. 19 of [107] shows the complete scheme of a setup with a submicron ion beam. To determine the diameter of the focused beam, we used the system briefly described in [53]. In this system, the ion beam is scanned across the blade tip synchronously with the oscilloscope beam, to the input of which a current signal from the blade is supplied.

The steepness of the current rise can determine the diameter of the beam. To calibrate the scale, a blade shift of a known value was used.

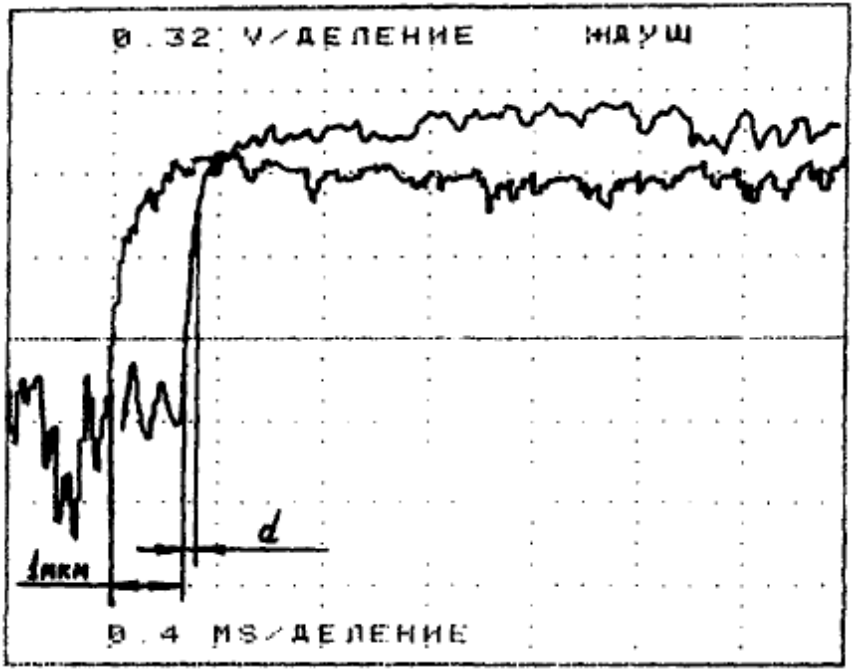

Fig. 24. Oscillograms of signals from a blade with a blade offset of 1 μm.
The estimated beam diameter is ~ 0.2 μm [53].

Oscillograms of signals from the blade when the blade is shifted by 1 μm are shown in Fig. 24. If we assume that the beam has a Gaussian distribution of the current density over the cross section, then the beam diameter d = 2σ should be measured between 16% and 84% of the maximum intensity. The estimated beam diameter is ~ 0.15-0.2 μm.
In Fig. 25 shows examples of microprocessing.

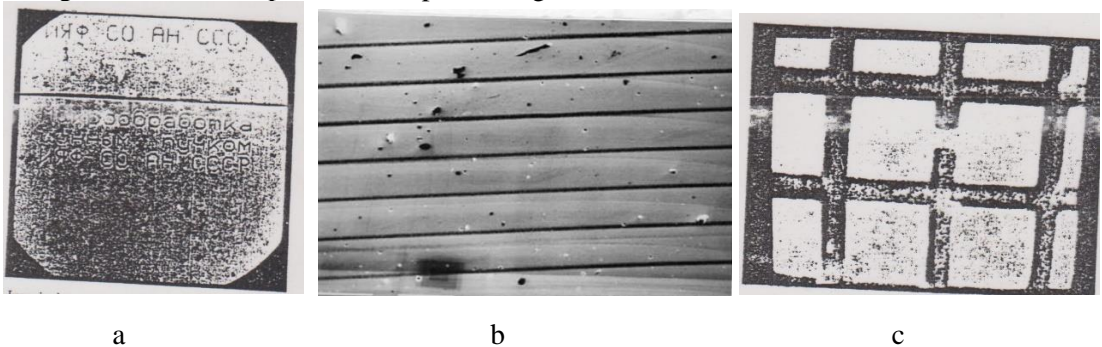

a                              b                                   c

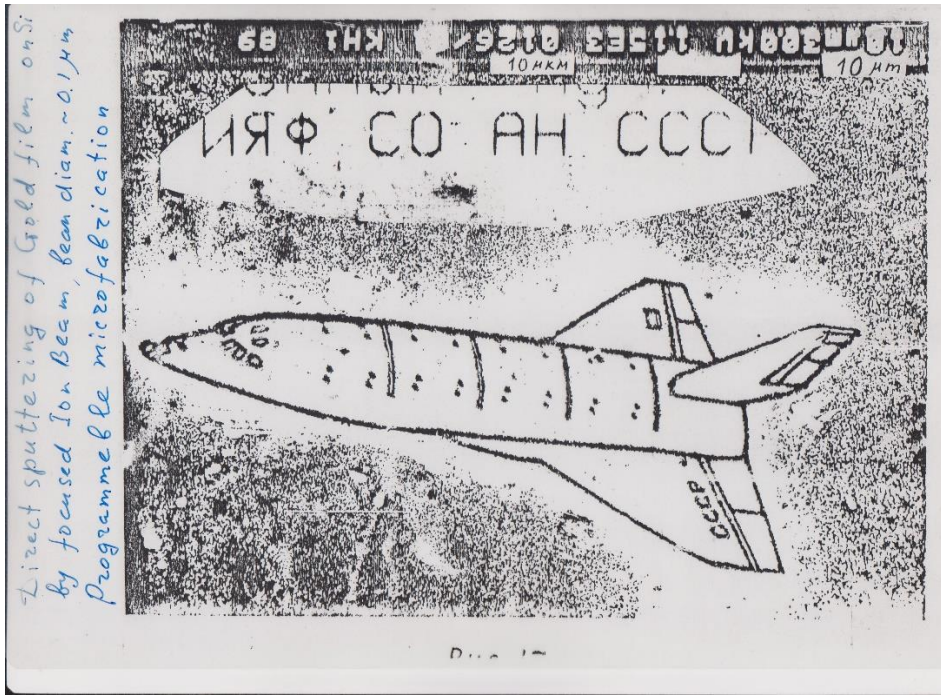

d

Fig. 25. Examples of microprocessing by a submicron ion beam [53].
a-fragment of the inscription made by the ion beam. The heights of the capital letters are 6 microns. b-raster drawn on a gold-plated silicon wafer, trace width 0.1 μm. c-grid cut by a focused ion beam (wire diameter of 6 μm). d-Figure of the spaceship t "Buran". The height of the capital letters is 6 microns.

The inscription is made on a silicon wafer with a sprayed layer of gold ~ 300 A. . The gold layer was sputtered in one pass of the beam. The image of the plate was obtained using a Raster Electron Microscope; the chemical contrast gold - silicon is visible. A special movement [122] was developed for the precision movement of the sample. The movement scheme is shown in Fig. 26. It consists of four piezoelectric plates 1 assembled in a square, four magnetic suction cups 2 and a ferromagnetic plate 3. Magnetic suction cups are magnetized alternately to the magnetic plate by passing current through the suction coil. Piezoelectric plates lengthen or shorten when voltage is applied, moving the slide along the ferromagnetic plate.
The maximum speed of movement is 4 mm / s, positioning accuracy is 0.05 mm, the positioning area is a circle with a diameter of 20 cm.
A photograph of the slide with the control module is shown in Fig. 27.

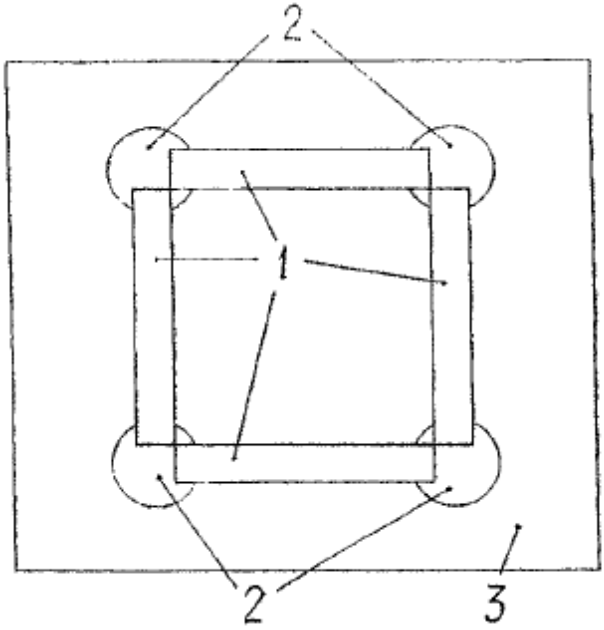

Fig. 26. Scheme of progress. 1-piezoelectric plates, 2-magnetic suction cups, 3-ferromagnetic plate.

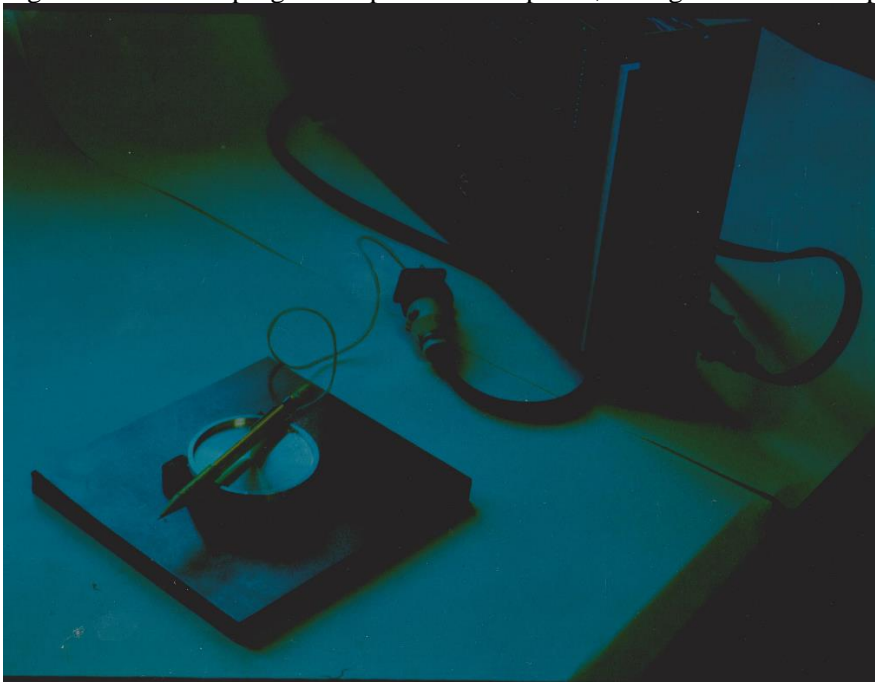

Fig. 27. Photo slider with a control module.

A diagram of an ion-optical column with a Wien filter for working with complex melts is shown in Fig. 28.

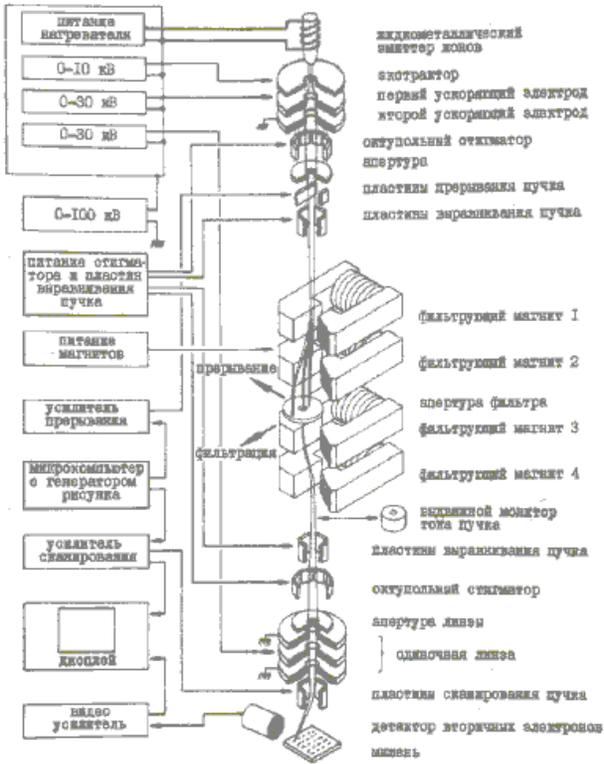

Fig. 28. Diagram of an ion-optical column with a Wien filter for working with complex melts MICROLAB-100 [123].

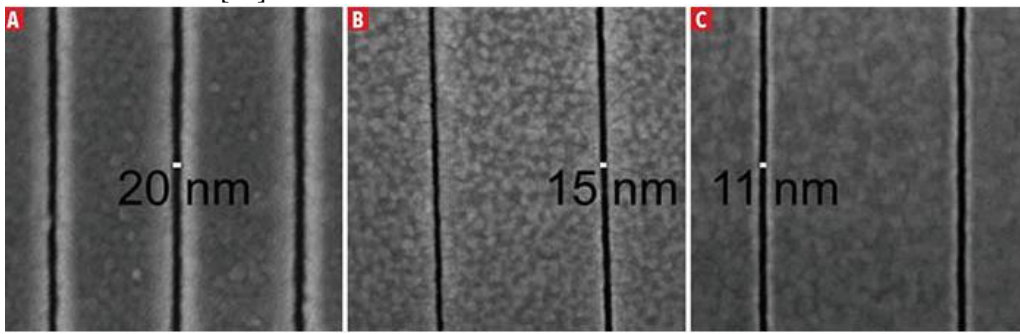

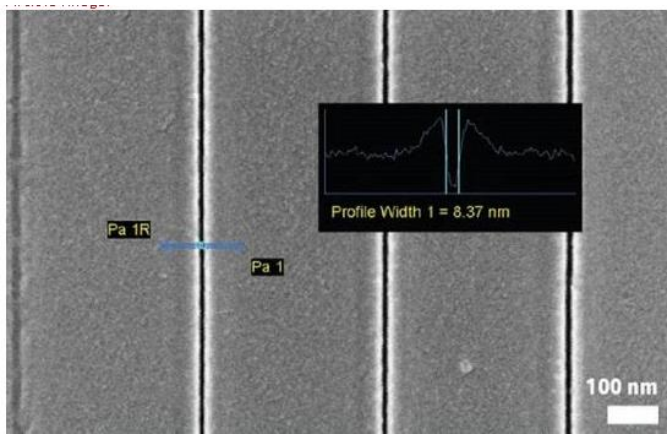

Fig. 29. SEM image of lines drawn by an ion beam on a gold film using (a) Au ++, (b) Si ++, and (c) Be ++ ions [124].

SEM image of lines drawn by an ion beam on a gold film using (a) Au ++, (b) Si ++, and (c) Be ++ ions are shown in Fig. 29.

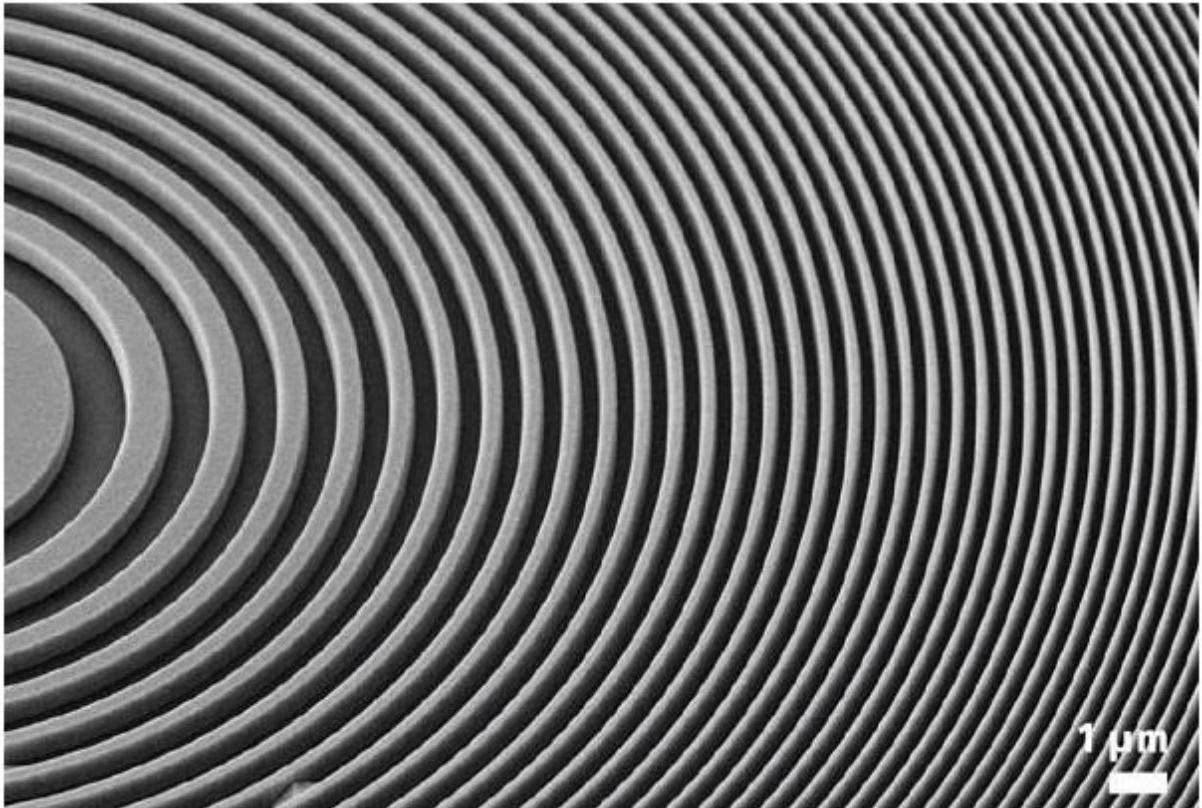

Fig. 30. Zone plate with a high aspect ratio made by submicron ion beam with high resolution [125].

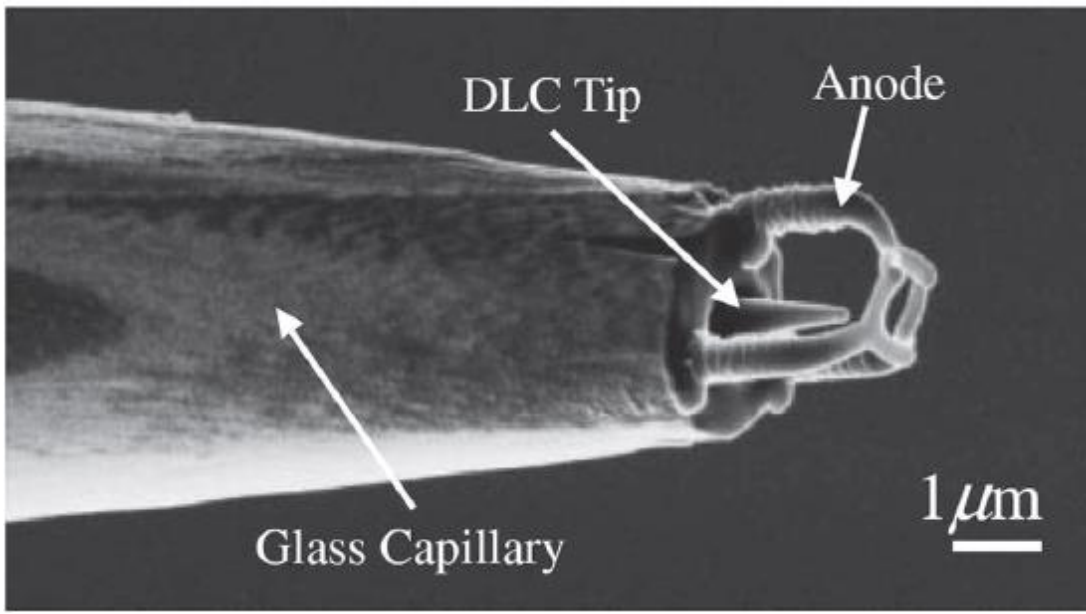

Fig. 31. Three-dimensional emitter and anode obtained by sputtering with a focused ion beam from a gas precursor [126].

A high-resolution image of a zone plate made by a high-resolution submicron ion beam is shown in Fig. 30.

An example of three-dimensional growth of the anode using sputtering by a focused ion beam from a gas precursor is shown in Fig. 31.

A number of ion-optical systems with EHD emitters are described in [127].

Review of work on ion sources performed at the Institute of Nuclear Physics. Budker until 1990 is presented in [128].

A commercially available installation with crossed electron and focused ion beams is shown in Fig. 32.

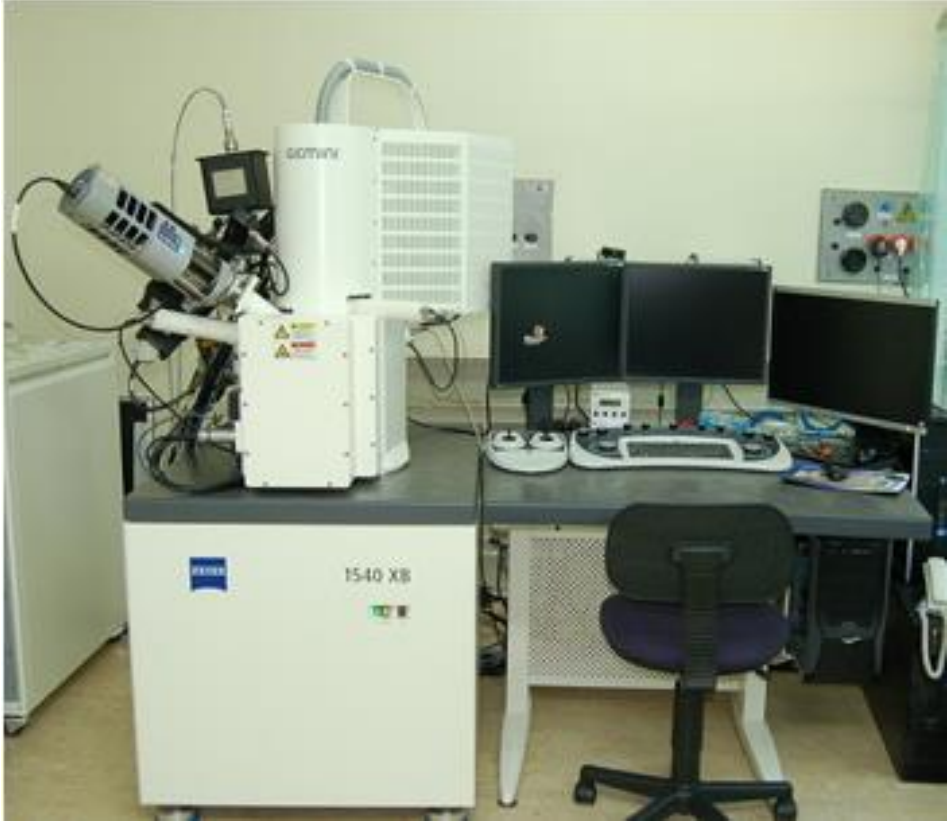

Fig. 32— CROSS BEAM 1540XB.
1. Manufacturer / Country - Zeiss, Germany.
2. Specifications:
o Features of the electronic gun:
♣ Working distance - 0-45 mm.
♣ Resolution - 1.1 nm @ 20 kV, 2.5 nm @ 1 kV.
♣ Magnification - 20x - 900kx.
♣ Beam current - 4 pA - 20 nA.
♣ Accelerating voltage - 0.1 - 30 kV.
♣ Cathode type - Thermal field emission type.
♣ Gun pressure <1 × 10-9 Torr. o Characteristics of the ion gun (Ga +):
♣ Working distance - 5 mm.
♣ Resolution - 5 nm is achievable.
♣ Magnification - 600x - 500kx.
♣ Beam current - 1 pA - 50 nA.
♣ Accelerating voltage - 3 - 30 kV.

♣ Type of cathode - Ga liquid metal ion source (LMIS).
♣ Gun pressure <1 × 10-9 Torr.
A review of ion sources based on low-temperature ionic liquids for aerospace applications, nanotechnology, and microprobe analysis is presented in [129]
In Fig. 33. An example of an ion source based on low-temperature ionic liquids is shown.

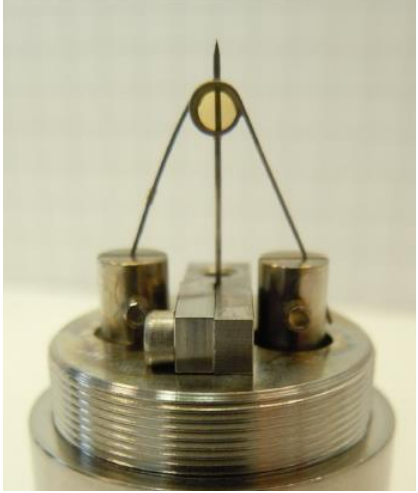

Fig. 33. An ion source based on low temperature ionic liquids. Ionic liquid
1-ethyl-3-methylimidazine tetrafluooborate (EMI-BF4) [130].

Figure 34 shows the dependence of the emission current on the voltage at the needle for an ion source based on low-temperature ionic liquids.

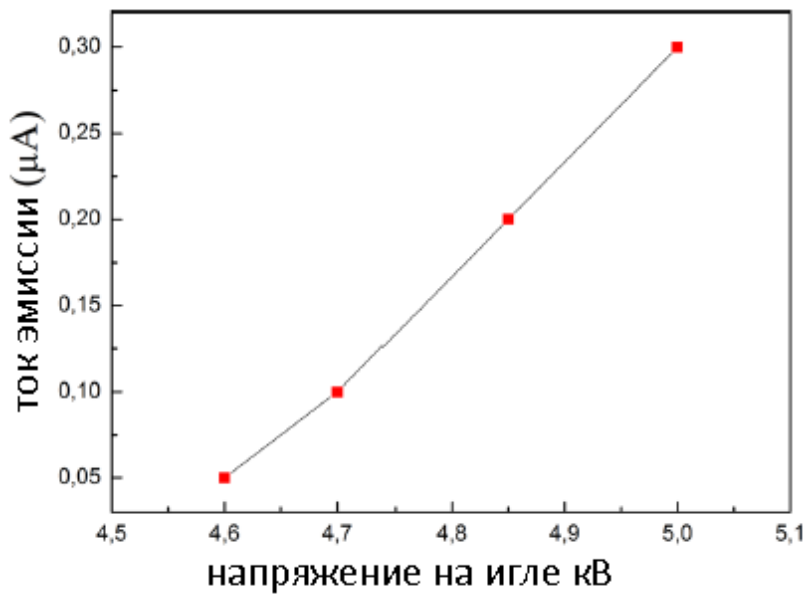

Fig. 34. Dependence of the emission current on the voltage at the needle for an ion source based on low-temperature ionic liquids.

FORMATION OF ION BEAMS WITH INCREASED INTENSITY

And finally, another possible application of EHD emitters is the formation of wide ion beams with increased intensity for accelerators and a number of other applications [53,[131],[132]].
The system for generating intense beams for accelerators is shown in Fig. 35.

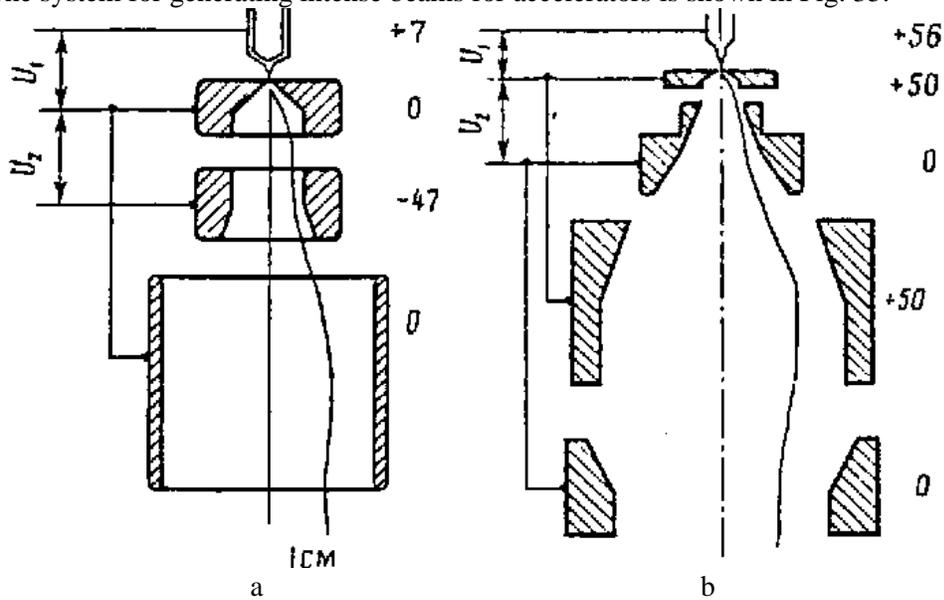

Fig. 35. System for the formation of intense ion beams for accelerators. a-ion beam with an energy of 7 keV, b-ion beam with an energy of 56 keV [53].

It consists of a pulling electrode and a single lens focusing a diverging beam. The distribution of current density in the ion beam after the lens is shown in Fig. 36. This source was installed on an ELV electrostatic accelerator and beams of gallium and indium ions with a current of up to 0.1 mA were obtained at an energy of 1.1 MeV.

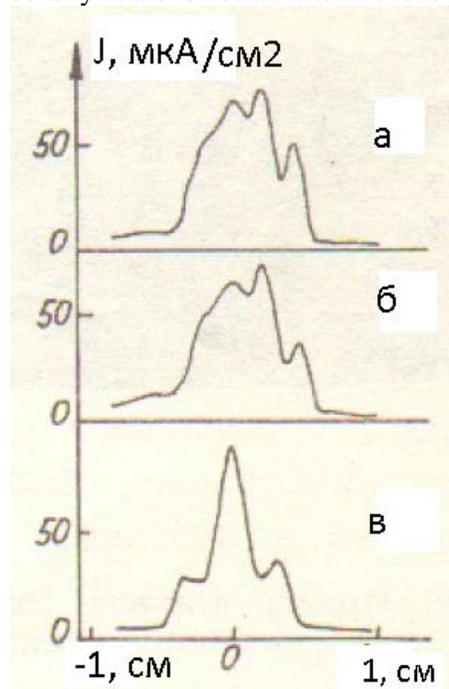

Fig. 36. The distribution of current density in the ion beam after the lens at various voltages $U_2$ on the middle electrode of the lens. a-- $U_2 = 47$ kV, b-$U_2 = 49$ kV, c-$U_2 = 50$ kV [53].

## CONCLUSION

The development of EHD emitters has become a fundamentally new method for producing ion beams with high brightness along with the development of three other fundamentally new methods for producing ions: the surface-plasma method for producing beams of negative ions [133,134,135], ECR sources of multiply charged ions [136], and electron-beam sources of multiply charged ions [137] .

Recent studies have clarified many previously obscure points in the functioning of electrohydrodynamic ion emitters. However, the available data is still fragmented and not linked into a complete consistent description. In particular, the reasons for the occurrence of specific emission oscillations with increasing ion current remain unclear. There are doubts about the stationarity of low emission currents. Some uncertainty remains in the identification of the main mechanisms of ion generation at high currents.

During the same years, significant progress has been achieved in improving the facilities for microprocessing by focused ion beams of EHD emitters. Interesting results were obtained in their technological applications. There were reports of successful commercial samples of installations with EHD emitters. It should be noted that the possibilities for increasing the parameters of focused beams of EHD emitters remain very significant. It can be hoped that physical studies of EHD emitters and the development of their effective applications in technologies will lead to new interesting results.

Despite tremendous advances in the technology of producing focused ion beams, modifications of vacuum systems of electron and ion microscopes to work in low vacuum, and automation of high-precision sample manipulators, systems with focused ion beams could not replace the photolithography process in the production of microcircuits. However, they have become an indispensable tool in the manufacture of masks for photolithography, correcting contacts on defective chips, as well as in the processes of debugging the production of experimental microcircuits.

I thank A. Shabalin and D. V. Kovalevsky for fruitful cooperation in the development of EHD emitters and systems based on them.